\documentclass[a4paper,11pt]{article}
\pdfoutput=1 

\usepackage{jcappub} 

\usepackage[T1]{fontenc} 
\usepackage{aas_macros}
\usepackage{booktabs}
\usepackage{multirow}
\usepackage{subcaption}
\usepackage{array}
\usepackage{comment}
\usepackage{parskip}
\usepackage{float}
\usepackage[dvipsnames]{xcolor}
\newcolumntype{M}[1]{>{\centering\arraybackslash}m{#1}}

\title{\centering\fontsize{22}{18}\selectfont Cosmological forecasts from the combination of Stage-IV photometric galaxy surveys and the magnification from forthcoming GW observatories}

\author[a,b]{Matteo Beltrame,}
\author[c,d,e,b]{Marco Bonici,}
\author[b]{Carmelita Carbone}

\affiliation[a]{Dipartimento di Fisica ``Aldo Pontremoli'', Universit\`{a} degli Studi di Milano, via Celoria 16, I-20133 Milano, Italy}
\affiliation[b]{INAF -- Istituto di Astrofisica Spaziale e Fisica cosmica di Milano (IASF-MI), via Alfonso Corti 12, I-20133 Milano, Italy}
\affiliation[c]{Waterloo Centre for Astrophysics, University of Waterloo, Waterloo, ON N2L 3G1, Canada}
\affiliation[d]{Department of Physics and Astronomy, University of Waterloo, Waterloo, ON N2L 3G1, Canada}
\affiliation[e]{Perimeter Institute for Theoretical Physics, 31 Caroline St. North, Waterloo, ON NL2 2Y5, Canada}

\emailAdd{matteo.beltrame1@studenti.unimi.it}
\emailAdd{mbonici@uwaterloo.ca}
\emailAdd{carmelita.carbone@inaf.it}

\abstract{In this work we have investigated the synergy between Stage-IV galaxy surveys and future GW observatories for constraining the underlying cosmological model of the Universe, focussing on photometric galaxy clustering, cosmic shear and GW magnification as cosmological probes. We have implemented a Fisher matrix approach for the evaluation of the full $6\times2$pt statistics composed by the angular power spectra of the single probes together with their combination. For our analysis, we have in particular considered dynamical dark energy and massive neutrino scenarios. We have found that the improvement to galaxy survey performance is below 1\%, in the case of $\ell^{\rm GW}_{\rm max}=100$ and a luminosity distance error of $\sigma_{d_L}/d_L=10\%$. However, when extending the analysis to $\ell^{\rm GW}_{\rm max}=1000$, we find that the GW magnification improves the galaxy survey performance on all the cosmological parameters, reducing their errors by $3\%$-$5\%$, when $\sigma_{d_L}/d_L=10\%$, and by $10\%$-$18\%$ when $\sigma_{d_L}/d_L=1\%$, especially for $M_\nu$, $w_0$ and $w_a$. However, here our analysis is unavoidably optimistic: a much more detailed and realistic approach will be needed, especially by including systematic effects. But we can conclude that, in the case of future gravitational wave observatories, the inclusion of the gravitational wave magnification can improve Stage-IV galaxy surveys performance on constraining the underlying cosmological model of the Universe.}


\arxivnumber{2403.03859}

\begin{document}
\maketitle
\flushbottom

\section*{Introduction}
\addcontentsline{toc}{section}{Introduction}
The observed cosmic acceleration of the Universe at late times and large scales is an established fact since 1998~\cite{Riess_1998, Perlmutter_1999}. 
According to the prevailing cosmological model, known as the $\Lambda$CDM model, this acceleration is driven by a cosmological constant $\Lambda$, postulating the existence of a dark energy (DE) component characterised by negative pressure and equation of state $w=-1$. 
This model delineates a cosmos comprised of three main constituents: firstly, there is ordinary baryonic matter, constituting 4.9\% of the total, which consists of baryons i.e. the typical atoms composing stars, gases, and galaxies. Secondly, we have cold dark matter, making up 26.8\% of the Universe, which is affected only by gravitational interaction. Finally, there is the cosmological constant accounting for 68.3\% of the cosmos (as per the findings of the Planck Collaboration~\cite{Planck2020}). This configuration aligns with the fundamental structure formation described by the Einstein's General Relativity (GR) equations in which the cosmological constant term is added. In this standard cosmological framework, the Universe is portrayed as a flat and expanding space-time, adhering to the principles of GR. It exhibits characteristics in agreement with various observational data, including the Cosmic Microwave Background (CMB), Supernovae observations, and Baryon Acoustic Oscillation (BAO) data.\\
However, the physical nature of the DE remains poorly understood. Extensions to this model involve changes in the properties of DE over time, as parameterised by an equation of state following the Chevallier-Polarski-Linder (CPL) model: $w_{\text{DE}}(z)=w_0+w_a\, z/(1+z)$~\cite{CHEVALLIER_2001, Linder}. Such changes in pressure affect not only the background expansion but also the evolution of the Large Scale Structure (LSS) of the Universe. Specifically, greater acceleration leads to slower growth of these structures. Although various models describing DE have been proposed~\cite{Peebles, Steinhardt}, the fundamental nature of DE remains a mystery. Precise measurements of $w(z)$ hold the promise of providing insights into its features~\cite{Gerardi:2019obr}.

Another enigma within our Universe pertains to the elusive mass of neutrinos. Until just a few years ago, the prevailing notion in the standard model of particle physics was that the three active neutrinos had no mass. However, in 1998, the Super-Kamiokande collaboration~\cite{Superkamio} unveiled the evidence of neutrino oscillations, implying that at least two of these subatomic particles possess mass. Additionally, the presence of massive relic neutrinos can influence both the growth of cosmic structures and the Universe expansion history. Their high thermal velocities result in free streaming, altering the epoch of matter-radiation equality and dampening the growth of structures at moderately nonlinear and small scales. Thus, the Universe can serve as a laboratory for constraining the scale of the neutrino mass. The significance of neutrinos in cosmology comes from the impact their masses exert on cosmological observables, particularly on LSS probes such as galaxy clustering and weak gravitational lensing. These effects hold substantial importance for two primary reasons. Firstly, the absolute scale of the neutrino mass remains still unknown. Consequently, the study of their effects on cosmological observables could play a crucial role in discerning this mass scale. Gravity responds to the total mass of neutrinos, denoted as $M_{\nu}\equiv\sum_i m_{\nu_i}$, and is relatively less sensitive to the differences in their mass values~\cite{LESGOURGUES_2006}. Hence, by investigating these cosmological effects, we gain a valuable avenue in determining the total neutrino mass. Secondly, achieving an accurate understanding of the impact of massive neutrinos on LSS is important to avoid systematic errors in the determination of fundamental cosmological parameters, including parameters related to DE, such as its density and equation of state.

On the one side, upcoming galaxy surveys are designed to explore both the equation of state of DE $w_{\text{DE}}$ and the mass scale of neutrinos through measurements of structure growth and distance scales. On the other side, forthcoming gravitational wave (GW) observatories promise to provide independent and complementary constraints on cosmological parameters. Cosmological investigations employ various observables, including e.g. galaxy clustering (GC), weak lensing (WL) and  GW magnification. At the GC level, data from LSS surveys are employed to infer cosmological parameters by taking advantage of the sensitivity of galaxy density fluctuations tracing the underlying dark matter (DM) density field.  At the WL level, the matter field generates \textit{cosmic shear} (CS), which refers to little deformations in galaxy shapes caused by the gravitational potential generated by intervening density perturbations that light traverses from the source to the observer. Combining these galaxy shape images with estimates of their redshifts enables us to measure the growth of cosmic structures and enhance our understanding of cosmological parameters. The notion of employing weak lensing of GW as a cosmological probe was initially discussed in~\cite{Cutler_2009}. Correlations found in the slight fluctuations of the GW amplitude offer a means to deduce characteristics of the matter distribution in the intervening space. Such observations introduce the prospect of employing GWs not only as a tool to explore the geometry of the Universe but also to investigate its constituents and the LSS in general~\cite{Takahashi_etal_2005, Camera_Nishizawa_2015, Congedo_Taylor_2019, Mpetha_etal_2023}. Exploiting GW weak lensing, as opposed to galaxies, to map the DM distribution of the Universe presents several advantages beyond being an innovative approach. It represents a remarkably clean measurement, sidestepping issues like intrinsic alignments or blending that frequently affect weak lensing observations. Additionally, as several works in the literature showed, cross-correlations of LSS probes in general improve the determination of cosmological parameters and the measurement of nuisance parameters~\cite{Camera_2016, EUCLID:2020jwq, Euclid:2021xmh, Euclid:2021qvm, Euclid:2022hdx, Euclid:2022qtk}. This can be extended also to the cross-correlation of GWs with LSS. Furthermore, GW detectors are capable of capturing mergers at considerably higher redshifts than galaxy lensing, extending up to approximately $z\sim 10-20$. Similar to the conventional GW source localisation case, the primary limitation in this analysis remains the determination of the GW source redshift. In this respect there are two possibilities to measure redshifts of the GW resolved events: on the one side, for the so-called \textit{bright sirens}, the detection of their electromagnetic counterpart can provide a measure of their redshift, on the other side, for the so-called \textit{dark sirens} without an electromagnetic counterpart, statistical techniques exploiting the cross-correlation of GW events with spectroscopic and/or photometric galaxy surveys may help in determining the GW source redshift~\cite{Menard:2013aaa}. By the 2040s, numerous other GW detectors, including the Einstein Telescope (ET)~\cite{Maggiore_2020JCAP}, Cosmic Explorer (CE)~\cite{CE}, and LISA~\cite{LISA}, whether ground-based or space-based, will either be actively collecting data or in the advanced stages of their development. The collective capabilities of these detectors will enable the exploration of a broader spectrum of GW frequencies, encompassing various source types, in comparison to the LIGO-Virgo-Kagra (LVK) network~\cite{Abbott_2020}. Additionally, these future detectors will exhibit significantly enhanced sensitivity, which will have a profound impact on the exploitation of GWs in both astrophysical and cosmological investigations. The availability of a large number of GW detections originating from binary systems will be invaluable for conducting statistical cosmological analyses.

This article presents forecasts on cosmological parameters based on the combination of CS, photometric galaxy clustering (GC$_{\text{ph}}$), GW magnification, and their cross-correlations. In order to constrain the parameters we exploited and implemented the Fisher matrix approach~\cite{fisher} using angular power spectra as probes computed in the Limber approximation~\cite{Limber_1953}.\\
This work is organised as follows. In section~\ref{sec:1} we present the validation pipeline of the code we have implemented for the Fisher matrix forecast and the methodology used in this work. In section~\ref{sec:2} we show the results and interpretations of the cosmological parameter forecasts from the combination and cross-correlation of Stage-IV galaxy surveys and GW future detectors. Finally, in the conclusions we briefly summarise the main results of this work.

\section{Probes and forecasting approach}
\label{sec:1}
In this work we present forecasts on cosmological parameters based on data from ongoing and forthcoming galaxy surveys combined with observations from future GW detectors. In this section, after a brief description of the considered experiments, we report the formalism followed in order to compute Fisher matrix forecasts and validate our results.

\subsection{Surveys}
\subsubsection{Stage-IV galaxy surveys}
We consider Stage-IV galaxy surveys such as \textit{Euclid}\footnote{http://www.euclid-ec.org/}~\cite{Euclid_RB} and the \textit{Roman} Space Telescope\footnote{https://roman.gsfc.nasa.gov/}~\cite{WFIRST} which will explore the expansion history of the Universe and the evolution of large scale cosmic structures by measuring shapes and redshifts of galaxies, covering more than $1/3$ of the sky, up to redshifts $z\sim 2$. They will be able to measure ten of millions of spectroscopic galaxy redshifts, which can be used for galaxy clustering measurements, and a few billions of photometric galaxy images, which can be used for weak lensing observations. Worth of mention is also the \textit{Rubin} Observatory\footnote{https://www.lsst.org/}, previously referred to as the Large Synoptic Survey Telescope (LSST)~\cite{LSST_2009}, which is an astronomical observatory currently under construction in Chile.
Joint analyses of their data, combined with observations of the CMB, will provide new insights to a broad spectrum of science cases, ranging from galaxy formation and evolution to DE equation of state and neutrino masses.

\subsubsection{Future GW detectors}
Among the forthcoming GW observatories there are third-generation (3G) ground-based detectors (see e.g. \cite{Punturo_2010, Maggiore_2020JCAP, Branchesi_2023JCAP}), space-based interferometers (see e.g. \cite{LISA, LISA_cosmology2023}), and eventually high angular resolution GW observatories~\cite{baker2019high}, which will tremendously improve GW observations with respect to current ones. In fact, while conventional electromagnetic telescopes have long excelled at obtaining high-precision spatial information, GW observations usually deviate from this: instead of offering precise positions of GW sources, they provide a wealth of alternative data about their intrinsic properties, such as masses, rotation rates, and distances, which are challenging to be obtained via electromagnetic observations. Improving the astrometric accuracy of GW source positions, which is a crucial limitations at the moment, holds the potential to associate them to a single galaxy or cluster.\\
Depending on their configuration, 3G GW observatories will be characterised by a luminosity distance error, $\sigma_{d_L}/d_L \sim 1\% - 10\%$, which for space-based interferometers is expected to achieve $\sigma_{d_L}/d_L \sim 0.2\% - 2\%$, while for arcminute resolution GW observatories $\sigma_{d_L}/d_L \sim 0.1\% - 1\%$. The latter detectors will introduce distinctive tools for addressing fundamental questions in cosmology and astrophysics. Moreover, achieving arcminute resolution in astrometric positions will guarantee the identification of the host galaxy for the majority of detected sources. This huge resolution will ensure an electromagnetic counterpart for all binary neutron star detections, leading to an increase in the number of standard sirens in a range from 10$^5$ to 10$^6$. This, in turn, will allow for precision measurements of cosmological parameters. Additionally, given that the GW source population serves as a biased tracer of the underlying DM distribution, correlations with galaxy surveys will offer new insights into cosmological structure formation and evolution. In this work we consider two cases: $\sigma_{d_L}/d_L=1 \%$ and $10 \%$, and show that assuming a $1 \%$ error for $d_L$ especially improves constraints on the DE equation of state and the total neutrino mass.

\subsection{Fiducial cosmology}
In the case of CS and GC$_{\text{ph}}$ we have followed the structure of~\cite{Euclid_2020} while for GW we refer to~\cite{Balaudo_2023}. We perform a tomographic approach, working in the harmonic space and considering as observables the number counts angular power spectra, defined in the section below and commonly denoted as $C(\ell)$, where $\ell$ represents the multipole moment. 

In this work we assume $\Omega_r=\Omega_k=0$ in the Hubble parameter $E(z)$ defined as:
\begin{equation}
    E(z)=\sqrt{\Omega_m(1+z)^3+\Omega_{\text{DE}}(1+z)^{3(1+w_0+w_a)}\exp\left(-3w_a\frac{z}{1+z}\right)},
\end{equation}
therefore neglecting the contribution of the radiation at the redshifts of interest, and considering the flat space case alone. Moreover, we account for the presence of massive neutrinos. 
Our goal is to make forecasts for different cosmological models such as $\nu\Lambda$CDM and $\nu w_0w_a$CDM scenarios, by adopting the following set of parameters:
\begin{itemize}
    \item the reduced Hubble constant $h$ defined as $h=H_0/(\,100$ km s$^{-1}$ Mpc$^{-1})$.
    \item the sum of the three active neutrino masses $M_{\nu}= \sum_i m_{\nu_i}=0.06$ eV, which defines the neutrino density parameter $\Omega_{\nu}=M_{\nu}$(eV)$/93.14\,h^2$.
    \item the various density parameters such as $\Omega_m=\Omega_c+\Omega_b+\Omega_{\nu}$ and $\Omega_{\text{DE}}=1-\Omega_m$ at the present time.
    \item the parameters of the DE equation of state, $w_0$ and $w_a$.
    \item the root mean squared (r.m.s.) of the matter linear density fluctuation $\sigma_8$, and the scalar spectral index $n_s$.
\end{itemize}
We summarise the parameters of interest and their fiducial values in table~\ref{tab:fiducial}:
\begin{table}[h]
\centering
\renewcommand\arraystretch{1.1}
\caption{Parameters varied in the forecast analysis, together with their values in the fiducial cosmology.}
\begin{tabular}{cccccccc}\hline
\label{tab:fiducial}
$\Omega_m$ & $\Omega_b$ & $w_0$ & $w_a$ & $h$ & $n_s$ & $\sigma_8$ & $M_{\nu}$ [eV]\\
\hline
 0.32 & 0.05 & $-$1 & 0 & 0.67 & 0.96 & 0.816 & 0.06 \\
\hline
\end{tabular}
\end{table}

\subsection{Angular power spectra}
From the definition of the angular power spectrum
\begin{equation}
\label{eqz:cl_def}
   \delta_{\ell\ell'}\delta_{mm'}C(\ell)=\langle a_{\ell m}a^*_{\ell'm'}\rangle, 
\end{equation}
i.e., $C(\ell)$'s are defined as the expectation value of the product
of the coefficients of the spherical harmonic decomposition of a quantity $f(\theta,\phi)$ on the
sphere and they quantify the amount of correlation between different angular scales in the observed field $f$. The coefficients $a_{\ell m}$ are defined as:
\begin{equation}
  a_{\ell m} = \int f(\theta, \phi) Y_{\ell m}^*(\theta, \phi) \sin\theta d\theta d\phi.
\end{equation}
In order to fix the parameters of our interest we compute the Fisher matrix (see Appendix~\ref{apndx_A}) using the $C(\ell)$'s as probes in the Limber approximation. In fact, while the full computation of the cosmic shear and galaxy power spectra is relatively laborious~\cite{Taylor_2018, LSSTDarkEnergyScience:2022lno}, due to spherical Bessel functions and several nested integrals, the Limber approximation~\cite{Limber_1953} simplifies the $C(\ell)$'s expression as
\begin{equation}
\label{eqz:Cl}
    C_{ij}^{\text{AB}}(\ell)=\frac{c}{H_0}\int\text{d}z\frac{W_i^{\text{A}}(z)W_j^{\text{B}}(z)}{E(z)r^2(z)}P_{mm}\left(\frac{\ell+1/2}{r(z)},z\right),
\end{equation}
where $r(z)$ is the comoving distance defined as $r(z)=\frac{1}{H_0}\int_0^z\frac{\text{d}z'}{E(z')}$, $W^{\rm A}_i(z)$ is the window function for a probe and $P_{mm}$ is the \textit{nonlinear} matter power spectrum.\\
We use \texttt{CLASS}~\cite{class} for computing the matter power spectrum $P_{mm}(k,z)$, which quantifies the statistical distribution and clustering of matter in the Universe at different scales.\\

In the presence of massive neutrinos, the matter power spectrum can be written with good accuracy as a sum of three contributions, that are the cold matter power spectrum $P_{cc}$, the neutrinos power spectrum $P_{\nu\nu}$ and the cross-power spectrum between cold matter and neutrinos $P_{c\nu} =\langle\delta_c\delta_{\nu}^*\rangle$, that is~\cite{Castorina2015}
\begin{equation}
    P_{mm}=(1-f_{\nu})^2P_{cc}+2f_{\nu}(1-f_{\nu})P_{c\nu}+f^2_{\nu}P_{\nu\nu}.
\end{equation}
The presence of neutrinos in the Universe has a significant impact on the total matter power spectrum, which is altered in a scale-dependent manner due to their free streaming~\cite{LESGOURGUES_2006}. Consequently, the matter power spectrum is a crucial quantity for constraining the total mass of neutrinos ($M_{\nu}$) using cosmological observations such as weak lensing and galaxy clustering. 

In the Limber approximation the relation between $k$ and $z$ is
\begin{equation}
    k_{\ell}(z)=\frac{\ell+1/2}{r(z)},
\end{equation}
so, we have interpolated the values of $P_{mm}(k_{\ell},z)$, starting from the previously computed $P_{mm}(k,z)$.

\begin{figure}
\begin{minipage}[b]{0.49\linewidth}
  \centering
 \includegraphics[scale=0.42]{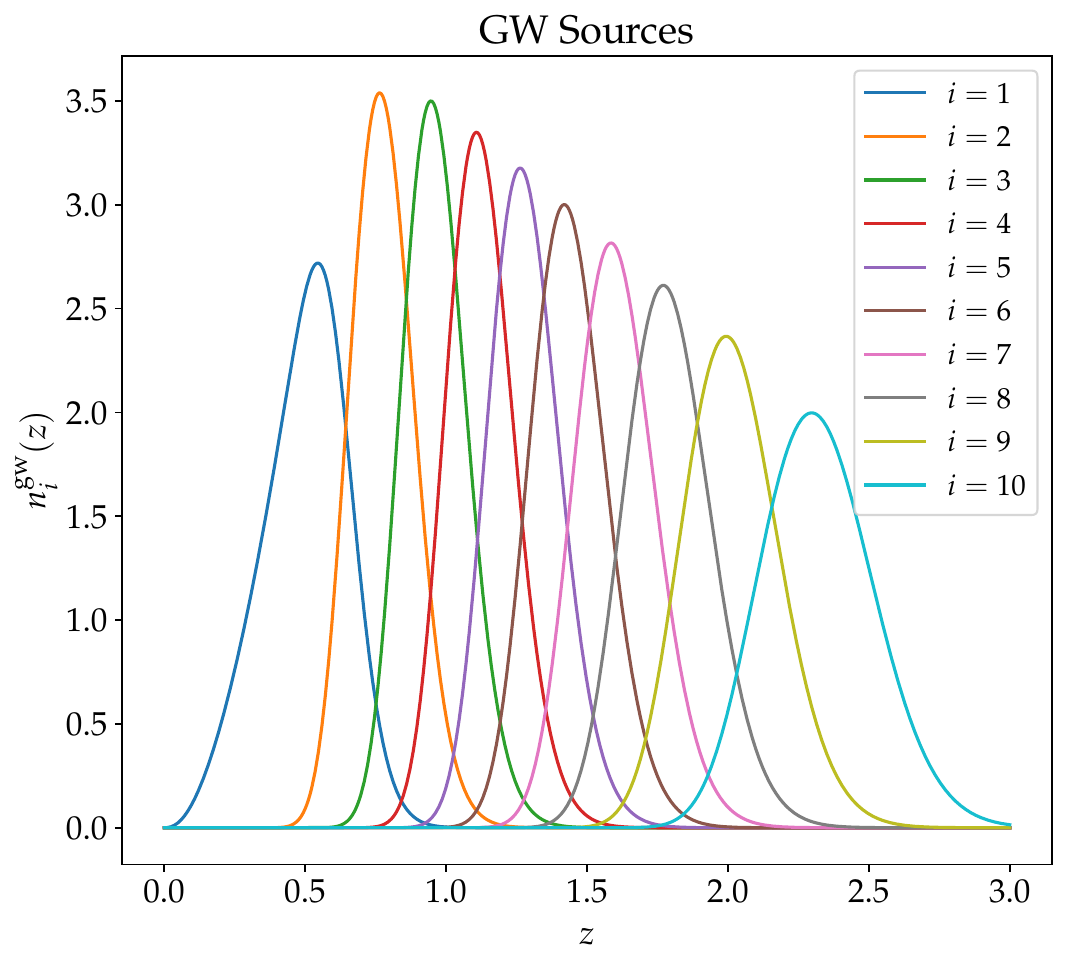}
\end{minipage}
\hfill
\begin{minipage}[b]{0.49\linewidth}
  \centering
   \includegraphics[scale=0.42]{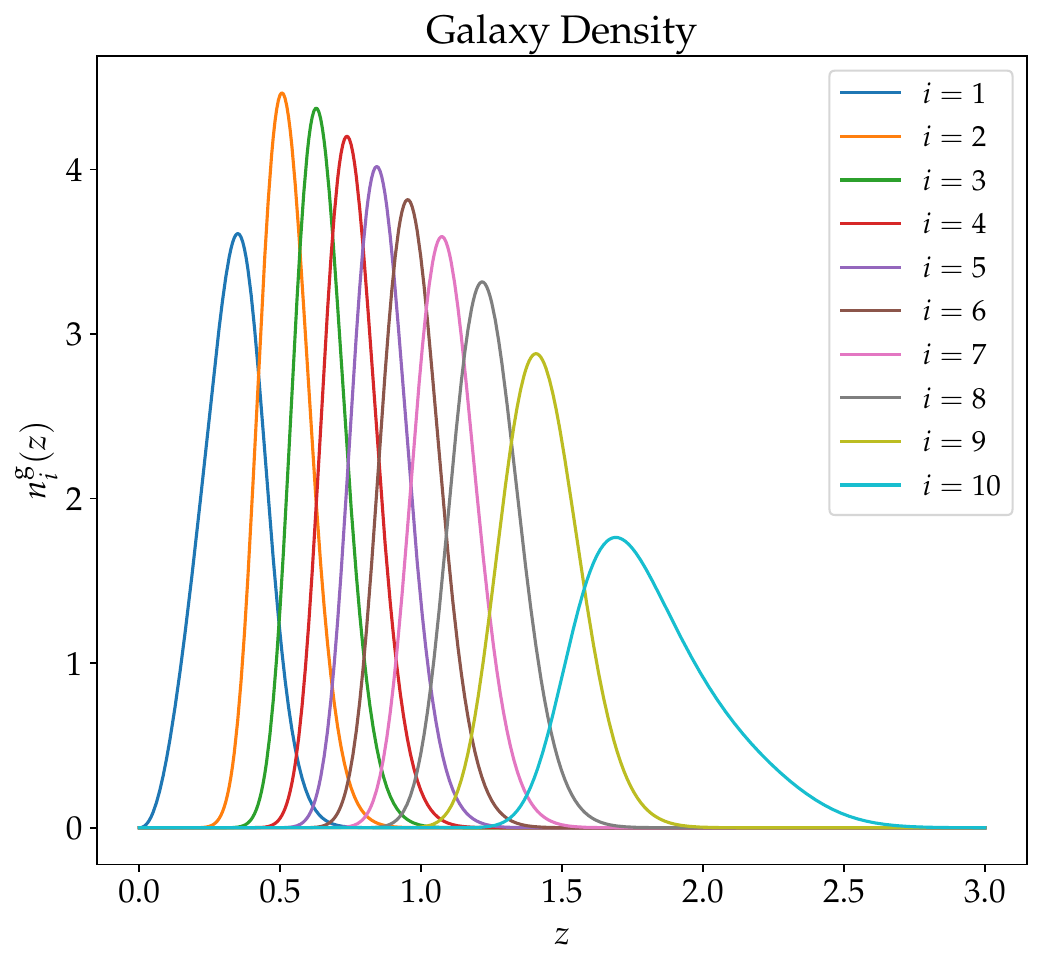}
\end{minipage}
\caption{Left: normalised distribution of GW sources. Right: Normalised number density of the observed galaxy distribution.}
\label{fig:galaxy}
\end{figure}

One of the relevant elements necessary to calculate eq.~\eqref{eqz:Cl} is the number density distribution $n_i(z)$ of the observed or host galaxies in the $i$-th bin. This redshift binning is known as \textit{tomography} and is required in order to achieve high-precision DE measurements. The process of redshift binning involves setting boundaries or ranges in redshift space within which galaxies are grouped. The exact binning scheme depends on the scientific objectives of the study and the available redshift measurements. One of the reasons why tomography is employed is to improve the cosmological constraints: dividing a sample into redshift bins can help mitigate the effects of cosmic variance, which arises from the random fluctuations in the large scale structure of the Universe~\cite{Hu_1999}.\\
For photometric estimates, which include errors in the measures, the distribution of galaxies is
\begin{equation}
\label{eqz:galaxy_den}
    n_i^a(z)=\frac{\int_{z_i^-}^{z_i^+}\text{d}z_pn(z)p_{\text{ph}}(z_p|z)}{\int_{z_{\text{min}}}^{z_{\text{max}}}\text{d}z\int_{z_i^-}^{z_i^+}\text{d}z_pn(z)p_{\text{ph}}(z_p|z)}, \quad \text{with}\quad a=(\rm g, gw)
\end{equation}
where $(z_i^-,z_i^+)$ are the edges of the $i$-th bin, while the true distribution $n(z)$ is
\begin{equation}
\label{eqz:n_i_z}
    n(z)\propto\left(\frac{z}{z_0}\right)^2\exp{\left[-\left(\frac{z}{z_0}\right)^{3/2}\right]},
\end{equation}
where $z_0=z_{\rm m}/\sqrt{2}$ for CS and GC$_{\text{ph}}$ ($n^{\rm g}_i(z)$), with $z_{\rm m}=0.9$ being the median redshift, and $z_0=1.5$ for GW sources ($n^{\rm gw}_i(z)$). Here, we assume the same galaxy distribution for the photometric shear and galaxy clustering surveys. With this configuration, one can set the edges of the 10 equi-populated bins for CS and GC$_{\text{ph}}$, which are
\begin{equation*}
    z^{\text{g}}_i = (0.001, 0.42, 0.56, 0.68, 0.79, 0.9, 1.02, 1.15, 1.32, 1.58, 2.5)
\end{equation*}
and for GW are
\begin{equation*}
    z^{\text{gw}}_i = (0.001, 0.631, 0.841, 1.013, 1.172, 1.329, 1.491, 1.667, 1.870, 2.124, 2.50)\,,
\end{equation*}
where $z_i^-=z_i$ and $z_i^+=z_{i+1}$. From eq.~\eqref{eqz:galaxy_den} one can see that the number density $n(z)$ is convolved in with the probability distribution function $p_{\text{ph}}(z_p|z)$, describing the probability
that a galaxy with redshift $z$ has a measured redshift $z_p$. A parameterisation for this quantity is
given by
\begin{equation}
    \begin{aligned}
    p_{\text{ph}}(z_p|z)=&\frac{1-f_{\text{out}}}{\sqrt{2\pi}\sigma_\text{b}(1+z)}\exp\left\{-\frac{1}{2}\left(\frac{z-\text{c}_{\rm b}z_p-z_\text{b}}{\sigma_\text{b}(1+z)}\right)^2\right\}+\\
    +&\frac{f_{\text{out}}}{\sqrt{2\pi}\sigma_\text{o}(1+z)}\exp\left\{-\frac{1}{2}\left(\frac{z-c_\text{o}z_p-z_\text{o}}{\sigma_\text{o}(1+z)}\right)^2\right\}
\end{aligned}
\end{equation}
By modifying the parameters of this function, one can mimic different cases of interest. Our choice is summarised in table~\ref{tab:probability}. We have fixed these quantities in the Fisher matrix estimate and do not explore their impact on the forecast.
\begin{table}[h]
\centering
\renewcommand\arraystretch{1.1}
\caption{Parameters to describe the photometric redshift distribution of sources.}
\label{tab:probability}
\begin{tabular}{ccccccc}\hline
$c_\text{b}$ & $z_\text{b}$ & $\sigma_\text{b}$ & $c_\text{o}$ & $z_\text{o}$ & $\sigma_\text{o}$ & $f_{\text{out}}$ \\
\hline
 1.0 & 0.0 & 0.05 & 1.0 & 0.1 & 0.05 & 0.1 \\
\hline
\end{tabular}
\end{table}
\\
 In figure~\ref{fig:galaxy} are shown the results of the implementation for eq.~\eqref{eqz:galaxy_den} for the different probes.

\subsubsection{Cosmic Shear}
\begin{figure}
\begin{minipage}{\linewidth}
  \centering
 \includegraphics[width=1.0\textwidth]{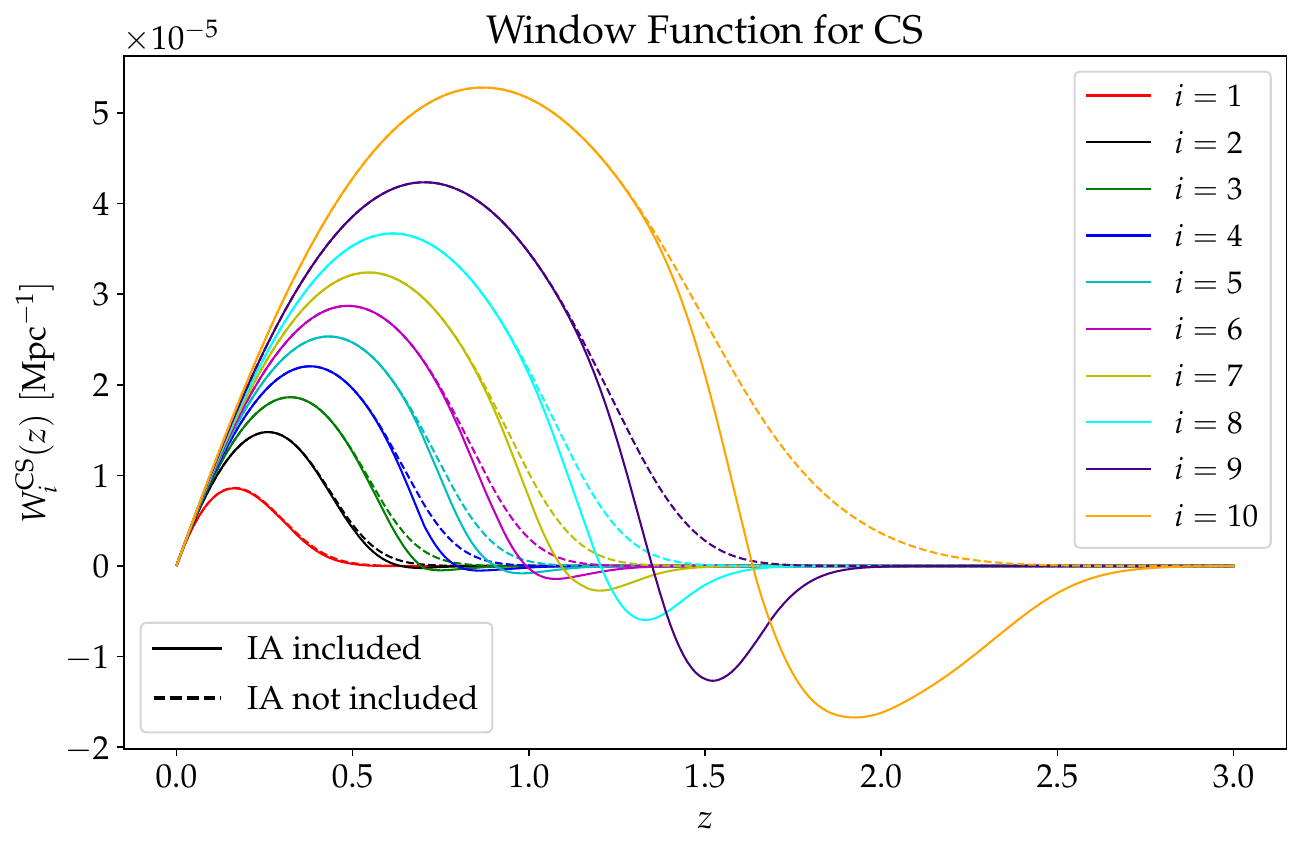}
\caption{Window function for CS. Dashed lines includes the contribution of the IA while solid lines represent eq.~\eqref{eqz:lens}.}
\label{fig:window_L}
\end{minipage}
\vspace{1em}

\begin{minipage}[b]{0.49\linewidth}
  \centering
 \includegraphics[width=\textwidth]{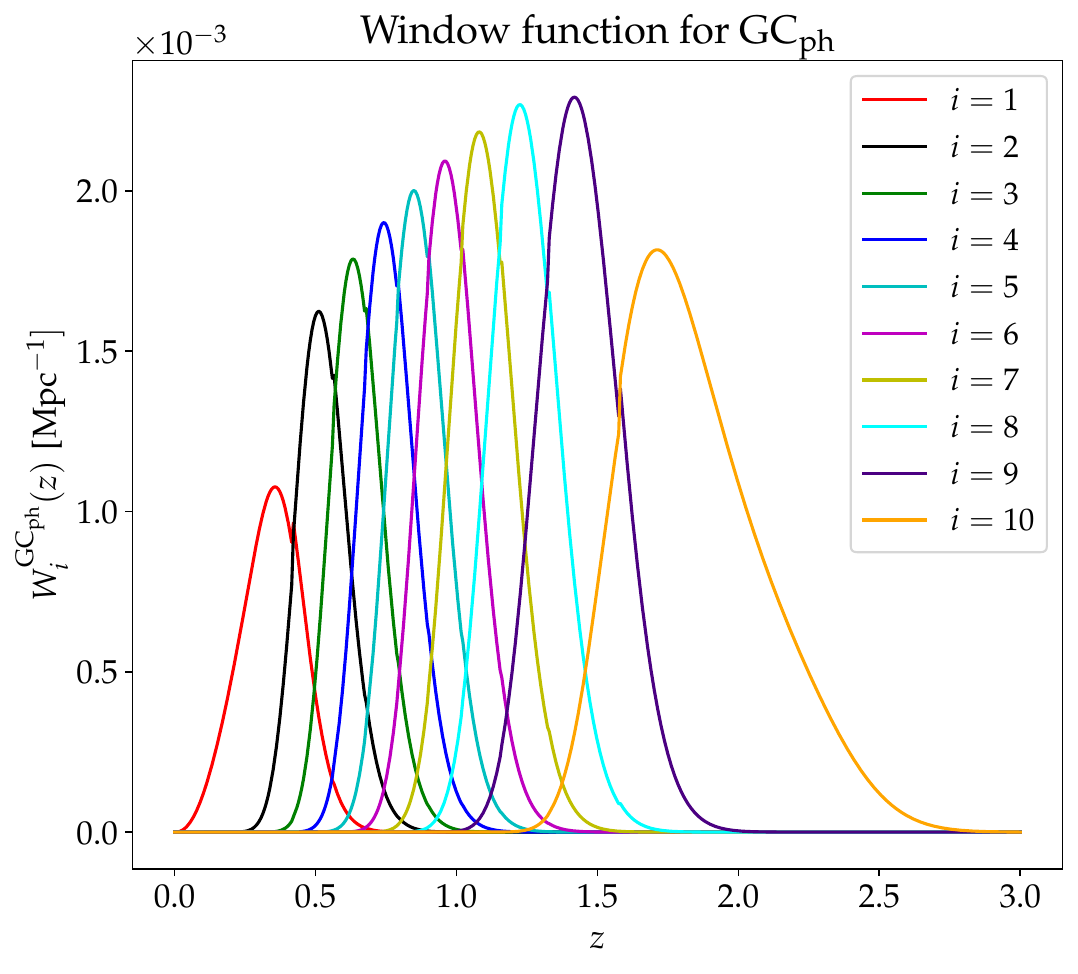}
\end{minipage}
\hfill
\begin{minipage}[b]{0.48\linewidth}
  \centering
   \includegraphics[width=\textwidth]{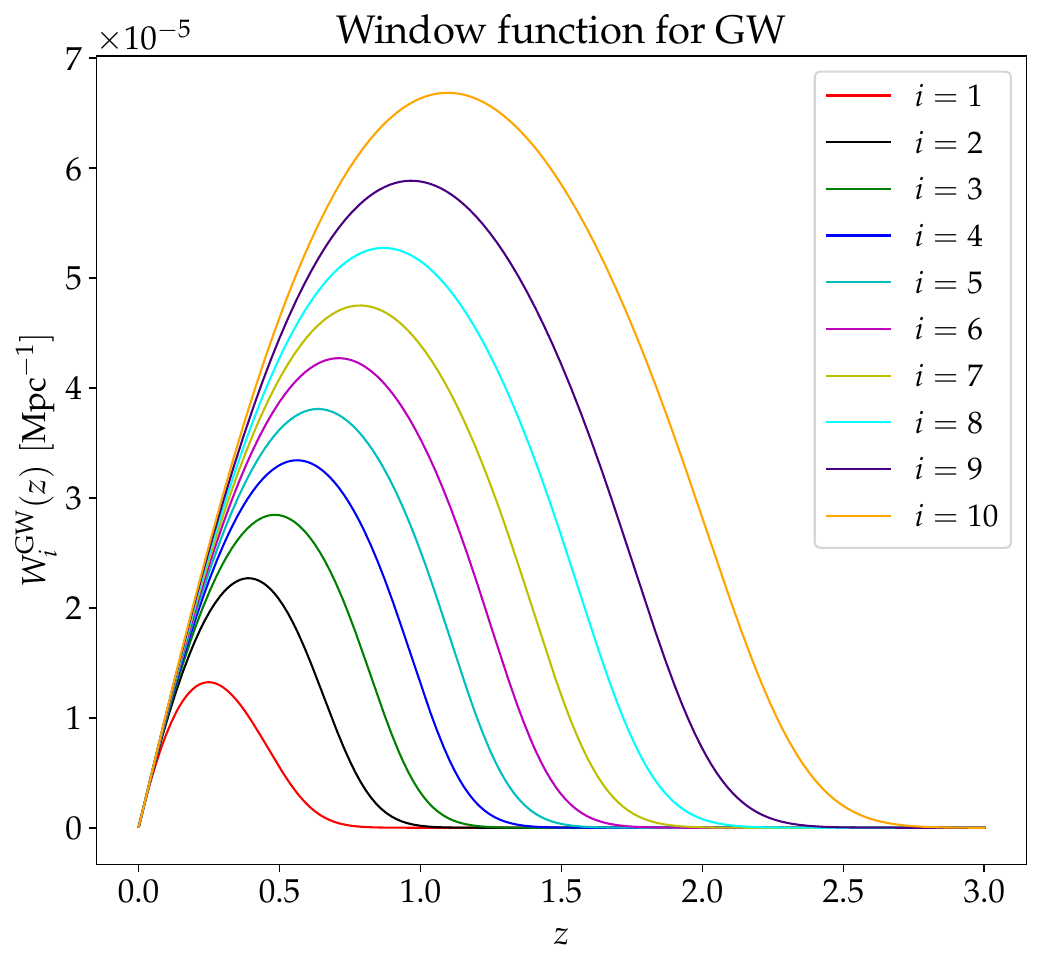}
\end{minipage}
\caption{Left: window function for GC$_{\text{ph}}$. Right: window function for the GW magnification.}
\label{fig:window_GW_GC}
\end{figure}

We want to evaluate eq.~\eqref{eqz:Cl} considering the cosmic shear as probe. To do so, one shall define the lensing window function as
\begin{equation}
\label{eqz:lens}
    W_i^{\gamma}(z)=\frac{3}{2}\left(\frac{H_0}{c}\right)^2\Omega_{m,0}(1+z)r(z)\int_z^{z_{\text{max}}}\text{d}z'n^{\rm g}_i(z')\left(1-\frac{\tilde{r}(z)}{\tilde{r}(z')}\right).
\end{equation}
However tidal interactions during the formation of galaxies and other processes may induce a preferred intrinsically correlated orientation of galaxy shapes, affecting the two-point shear statistics; tidal alignment refers to the arrangement of galaxy shapes or orientations induced by the gravitational tidal forces from nearby matter distributions. These tidal forces can be produced by both large scale structures and the gravitational potential of other galaxies. This implies that the intrinsic alignment (IA) is an important consideration when studying CS because it can contaminate the measurements of the cosmic shear power spectrum. CS is sensitive to the total gravitational lensing signal, which includes both the lensing effect from large scale structure and the IA signal. The window function for IA can be conveniently written as
\begin{equation}
\label{eqz:IA}
    W_i^{\text{IA}}(z)=-\mathcal{A}_{\text{IA}}\mathcal{C}_{\text{IA}}\Omega_m\frac{\mathcal{F}_{\text{IA}}(z)}{D(z)}\left(\frac{H_0}{c}\right)n^{\rm g}_i(z)E(z).
\end{equation}
where, $D(z)$ is the linear growth factor, while the function $\mathcal{F}_{\rm IA}(z)$ reads
\begin{equation}
    \mathcal{F}_{\text{IA}}(z)=(1+z)^{\eta_{\text{IA}}}\left(\left<L(z)\right>/\left<L_*(z)\right>\right)^{\beta_{\text{IA}}}
\end{equation}
The parameters $\eta_{\rm IA}$, $\beta_{\text{IA}}$ and $\mathcal{A}_{\text{IA}}$ are free parameters of the model. Here, we fix them to the following fiducial values: ($\eta_{\text{IA}}$, $\beta_{\text{IA}}$, $\mathcal{A}_{\text{IA}}) = (2.71, -0.41, 1.72)$, while $\mathcal{C}_{\text{IA}} = 0.0134$ remains fixed in the Fisher analysis. $L(z)$ and $L_*(z)$ are the redshift-dependent mean and the characteristic luminosity of source galaxies.\footnote{Their values were taken from \nolinkurl{http://pc.cd/0wJotalK.}. A file called \texttt{scaledmeanlum-E2Sa.dat} inside the WL folder contains the required data}
The CS window function is the sum of eqs.~\eqref{eqz:lens} and \eqref{eqz:IA}:
\begin{equation}
    W_i^{\text{CS}}=W_i^{\gamma}+W_i^{\text{IA}}.
\end{equation}
The plot of the implemented window function $W_i^{\text{CS}}$ is shown in figure~\ref{fig:window_L}.
The uncorrelated part of the intrinsic (unlensed) ellipticity field acts as a shot noise term in the observed power spectrum. This is non-zero for auto-correlation (intra-bin) power spectra, but is zero for cross-correlation (inter-bin) power spectra, because ellipticities of galaxies at different redshifts should be uncorrelated. This term can be written as
\begin{equation}
    N_{ij}^{\text{CS}}=\frac{\sigma_{\text{CS}}^2}{\bar{n}_{{\rm g},i}}\delta_{ij}^K
\end{equation}
where $\bar{n}_{{\rm g},i}$ is the galaxy surface density per bin, and has to be consistently expressed in inverse steradians; the surface density of galaxies is $\bar{n}_{\rm g}=30$ arcmin$^{-2}$ and one can notice that the number density in each bin is simply $\bar{n}_{\rm g}/10$, since 10 equi-populated bins are used. $\delta_{ij}^K$ is the Kronecker delta and $\sigma_{\text{CS}}^2=(0.3)^2$ is the variance of the observed ellipticities.\\
Now, it can be finally expressed the final form of the observed tomographic cosmic shear angular power spectrum in the flat-sky and Limber approximation:
\begin{equation}
\label{eqz:Cl_L}
    \tilde{C}_{ij}^{\text{CSCS}}(\ell)=C_{ij}^{\text{CSCS}}(\ell) + N_{ij}^{\text{CS}}.
\end{equation}
The left image in the first row of figure~\ref{fig:Cls} shows the plot of eq.~\eqref{eqz:Cl_L} that we have implemented without taking account of the shot noise $N_{ij}^{\text{CS}}$.

\subsubsection{Photometric galaxy clustering}
The other probe we want to take into account is the photometric galaxy clustering. In order to evaluate the angular power spectra the same procedures as for CS can be followed. The window function reads
\begin{equation}
\label{eqz:window_G}
    W_i^{\text{GC}_{\text{ph}}}(z)=b_i(z)n^{\rm g}_i(z)\frac{H(z)}{c},
\end{equation}
where $b_i(z)$ is the galaxy bias in tomography bin $i$. Here, a constant value in each redshift bin has been assumed and their fiducial values are:
\begin{equation}
    b_i(z)=\sqrt{1+\bar{z}_i},
\end{equation}
where $\bar{z}_i$ is the mean redshift of each bin. The plot of eq.~\eqref{eqz:window_G} is shown in the left side of figure~\ref{fig:window_GW_GC}.\\
In this case the shot noise for GC$_{\text{ph}}$ is $N_{ij}^{\text{GC}_{\text{ph}}}=1/\bar{n}_{\text{g},i}\delta_{ij}^K$ and the observed angular power spectra for GC$_{\text{ph}}$ is simply
\begin{equation}
\label{eqz:Cl_G}
    \tilde{C}_{ij}^{\text{GC}_{\text{ph}}{\text{GC}_{\text{ph}}}}(\ell)=C_{ij}^{\text{GC}_{\text{ph}}{\text{GC}_{\text{ph}}}}(\ell) + N_{ij}^{\text{GC}_{\text{ph}}}.
\end{equation}
In the right image of the first row of figure~\ref{fig:Cls} it is shown the plot of eq.~\eqref{eqz:Cl_G} that we have implemented without taking account of the shot noise $N_{ij}^{\text{GC}_{\text{ph}}}$.

\subsubsection{Gravitational wave magnification}
The last probe we are going to consider is the weak lensing of GW which refers to the distortion of the wavefront caused by the gravitational field of intervening matter. This effect is analogous to the gravitational lensing of light. When a GW passes through a region with matter, such as a galaxy cluster, the matter's gravitational field can act as a lens, causing the wavefront to be bent and distorted.\\
The weak lensing effect on GW can be described using the formalism of gravitational lensing\footnote{See \url{https://oguri.github.io/lectures/2020kek/note_2020kek.pdf} for full details.}, which involves the lensing potential $\psi$ related to the lensing convergence $\kappa$ by
\begin{equation}
  \kappa=\frac{1}{2}\nabla^2_{\boldsymbol{\theta}}\psi,
\end{equation}
where $\nabla^2_{\boldsymbol{\theta}}$ is the transverse gradient. The lensing convergence for a GW, $\kappa_{\rm gw}$, is related to the observed strain $h_{\text{obs}}$ and the intrinsic strain $h$ of the GW by the equation~\cite{Bertacca_2018, Takahashi_2006}:
\begin{equation}
    h_{\text{obs}}(t_o) = (1+\kappa_{\text{gw}})h(t_s).
\end{equation}
where $t_s$ and $t_o$ are the times measured by the clock of the source and the observer, respectively. This equation indicates that the observed strain of the GW is modified by the lensing convergence. The lensing convergence can amplify or diminish the observed strain, depending on the distribution of matter and the geometry of the lensing system. We know that photons and GW encounter structure which induces scale-dependent corrections to their luminosity distance. In these corrections not only weak lensing is considered but many other effects are included such as Doppler, Sachs-Wolfe, Integrated Sachs-Wolfe, Shapiro time delays and volume effects. The observed luminosity distance $d^{\text{obs}}_L$ becomes then a function of redshift and angular direction~\cite{Mukherjee_2020, Balaudo_2023, Garoffolo_2020} and can be written as:
\begin{equation}
\label{eqz:d_l_WL}
    d^{\text{obs}}_L(z,\hat{n}) = d^{\text{true}}_L(z)+ \Delta d_L(z, \hat{n}).
\end{equation}
where $d^{\text{true}}_L(z)$ is the real luminosity distance of the source. Here, WL is the dominant correction which it can also be a valuable signal to be exploited, rather than only a source of error. Since WL is an integrated effect it builds up along the propagation. By reaching $z\simeq2.5$ and focusing on high redshift GW sources, we can approximate $\Delta d_L/d^{\text{true}}_L\simeq\kappa_{\text{gw}}$. Therefore we can construct an estimator to get the WL convergence of the GW as~\cite{Mpetha_2023}
\begin{equation}
    \hat{\kappa}_{\text{gw}}(z,\hat{n})\equiv 1-\frac{d^{\text{obs}}_L(z,\hat{n})}{d^{\text{true}}_L}.
\end{equation}
\noindent This estimator can be biased by three factors: the experimental error on the $d^{\text{obs}}_L(z, \hat{n})$ measurement $\epsilon_{\text{gw}}$, the error on the source redshift $\epsilon_z$, or the choice of the wrong fiducial model $\epsilon_c$. By taking into account these bias sources, the convergence is modified as follows~\cite{Balaudo_2023}
\begin{equation}
\label{eqz:estim}
    \hat{\kappa}_{\text{gw}}(z,\hat{n})=\frac{d^{\text{true}}_L(1-\kappa_{\text{gw}})+\epsilon_{\text{gw}}}{d^{\text{true}}_L(1+\epsilon_z+\epsilon_c)}\simeq \kappa_{\text{gw}}-\epsilon_{\text{gw}}+\epsilon_z+\epsilon_c
\end{equation}
where, assuming $\kappa_{\text{gw}}$ to be small and of the same order as the $\epsilon_i$, we linearised to first order in $\kappa_{\text{gw}}$ and the $\epsilon_i$. This is valid for the GW sources with electromagnetic counterparts, while for those GW sources without electromagnetic counterparts, the left side of eq.~\eqref{eqz:estim} is the estimator of the lensing signal, which produces a multiplicative bias in estimation of the lensing potential $\kappa_{\text{gw}}$.\\
Therefore we adopt eq.~\eqref{eqz:lens} for the window function of the GW convergence with a slight difference in the evaluation of $n_i^{\rm gw}(z)$, as already mentioned. After the evaluation of the window function, which is plotted in the right side of figure~\ref{fig:window_GW_GC}, the computation of the angular power spectra, plotted in the left side of the second row of figure~\ref{fig:Cls}, is straightforward:
\begin{equation}
    \tilde{C}_{ij}^{\text{GWGW}}(\ell)=C_{ij}^{\text{GWGW}}(\ell) + N_{ij}^{\text{GW}}(\ell).
\end{equation}
The shot noise for the lensed GW is modelled as follow~\cite{Balaudo_2023}:
\begin{equation}
    N_{ij}^{\text{GW}}(\ell)=\frac{1}{\bar{n}_{\text{gw},i}}\left(\frac{\sigma^2_{d_L}}{d_L^2}+\frac{\sigma^2_s}{d_L^2}\right)e^{\frac{\ell^2\theta^2_{\text{min}}}{8\ln 2}}\delta_{ij}.
\end{equation}
In the noise term $\sigma_{d_L}$ represents the average experimental error on the luminosity distance of the GW sources; $\sigma_{\text{s}} = (\partial d_L/\partial z)\sigma^z_{\text{gw}}$ is the contribution to the luminosity
distance error brought by the uncertainty on the merger redshift $\sigma^z_{\text{gw}}=\sigma_{\text{gw}}(1+z)$, with $\sigma_{\text{gw}}=0.05$; $\bar{n}_{\text{gw},i}$ can be written as the product between $N_{\text{gw}} /4\pi$ and the integral of the normalised source distribution over each bin $n_i^{\rm gw}(z)$, where $N_{\text{gw}}=10^6$ is the number of detected sources. Lastly, $\theta_{\text{min}}=\pi/\ell_{\text{max}}$ is the sky-localization area of the GW event, which also dictates the maximum available multipole for the analysis. Since the noise term is $z-$dependent, we chose to associate to $z$ the central value of the $i$th bin considered.

\subsubsection{Probe combination and cross-correlation}
\begin{figure}[!ht]
\begin{minipage}[b]{0.49\linewidth}
  \centering
 \includegraphics[scale=0.395]{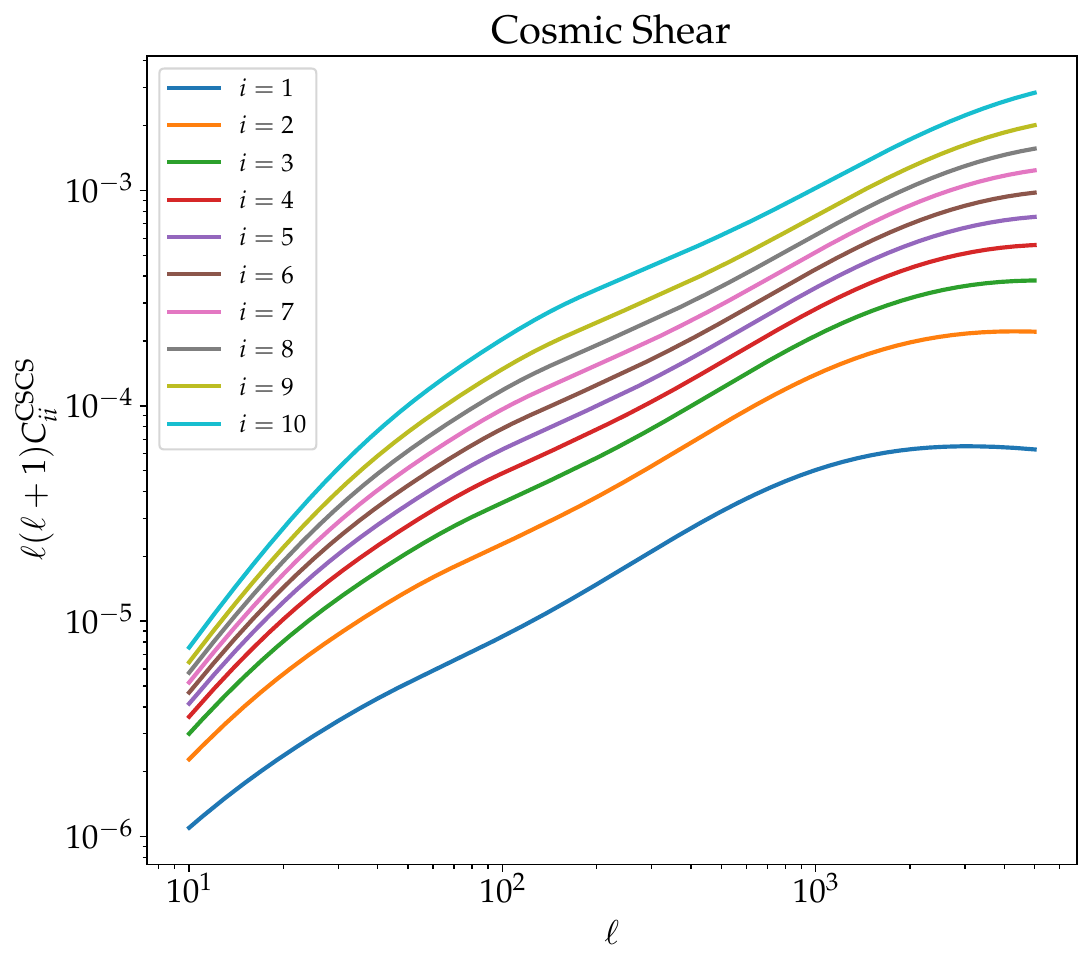}
\end{minipage}
\hfill
\begin{minipage}[b]{0.49\linewidth}
  \centering
   \includegraphics[scale=0.395]{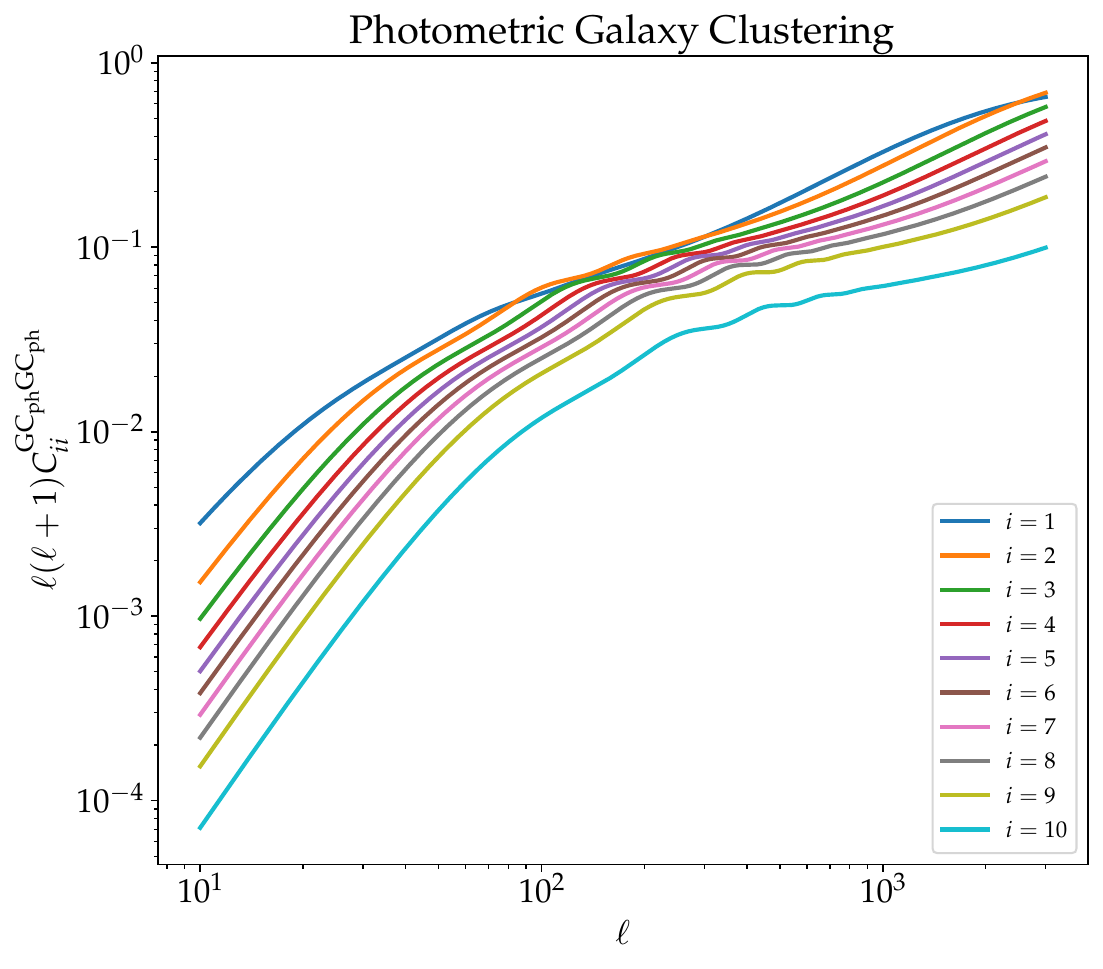}
\end{minipage}
\begin{minipage}[b]{0.49\linewidth}
  \centering
 \includegraphics[scale=0.395]{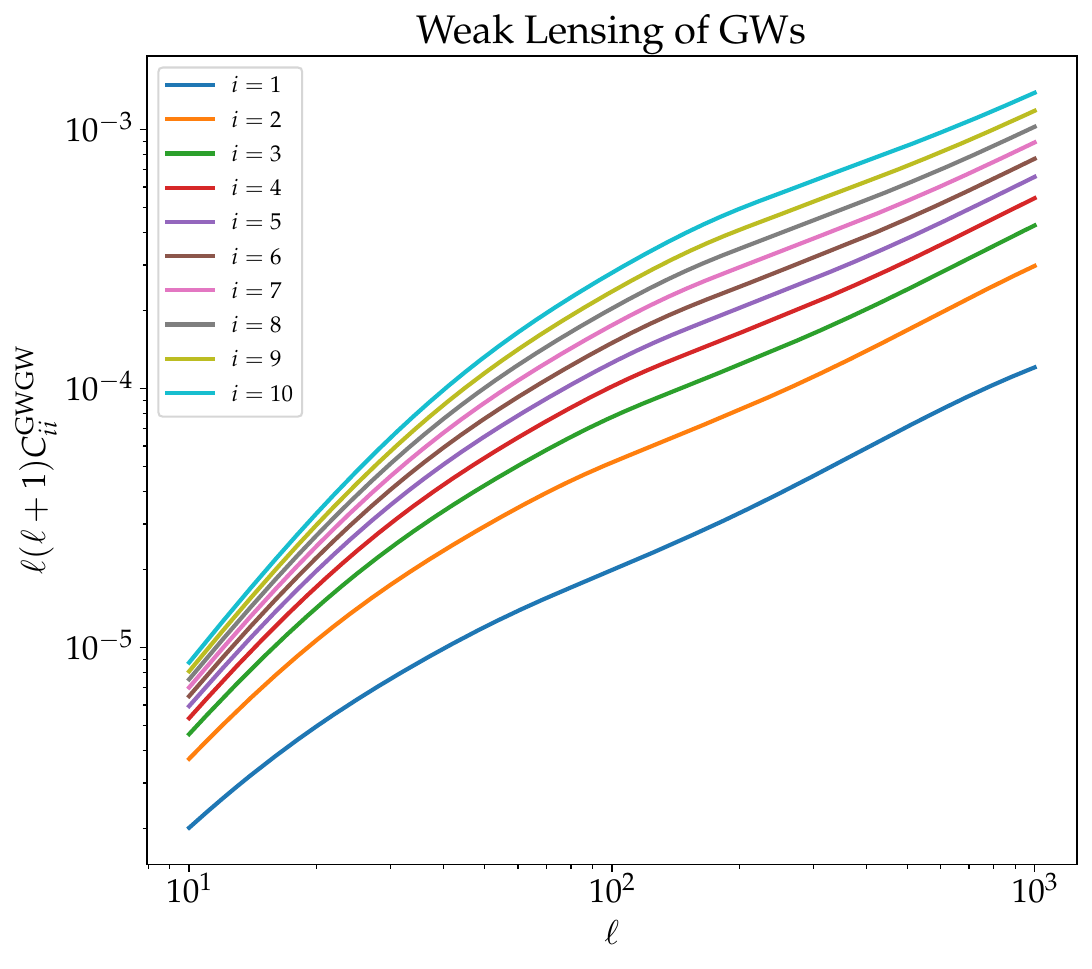}
\end{minipage}
\hfill
\begin{minipage}[b]{0.49\linewidth}
  \centering
   \includegraphics[scale=0.395]{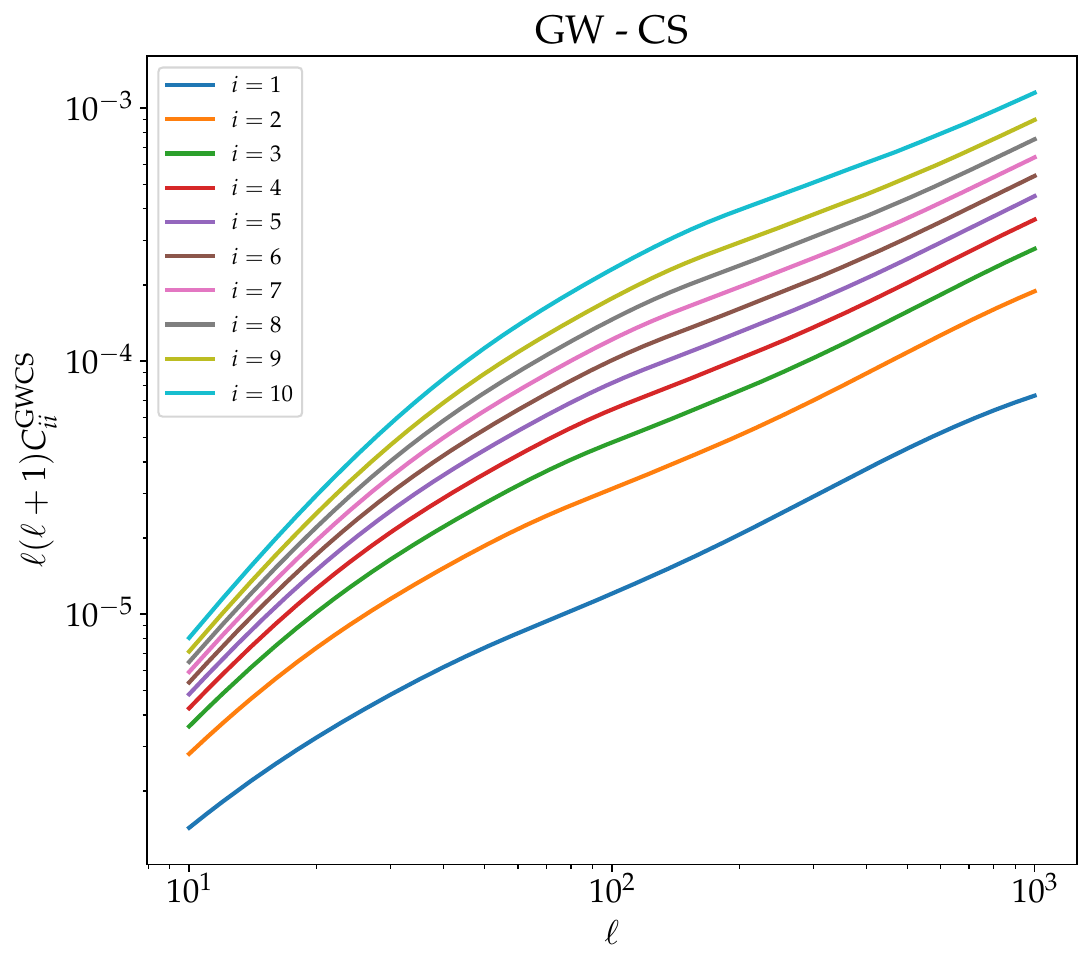}
\end{minipage}
\begin{minipage}[b]{0.49\linewidth}
  \centering
 \includegraphics[scale=0.395]{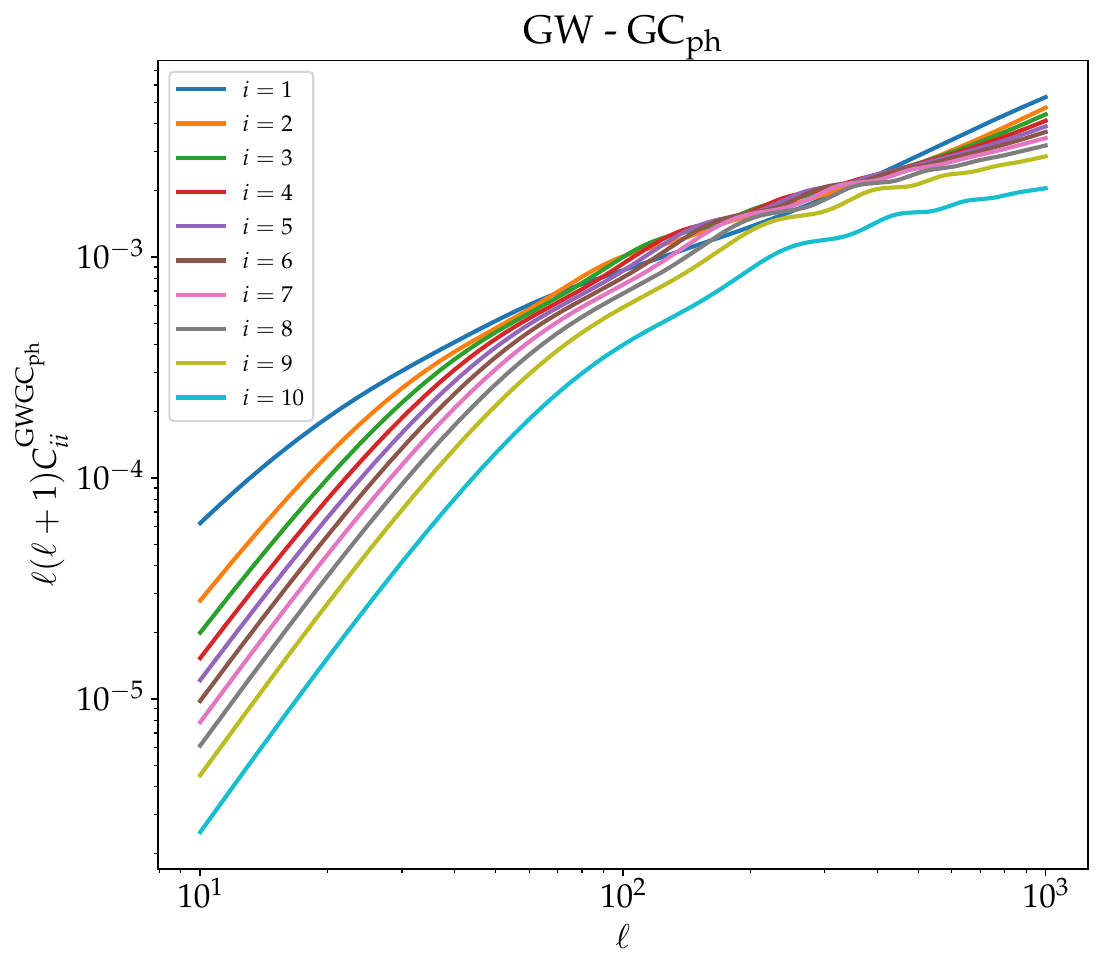}
\end{minipage}
\hfill
\begin{minipage}[b]{0.49\linewidth}
  \centering
   \includegraphics[scale=0.395]{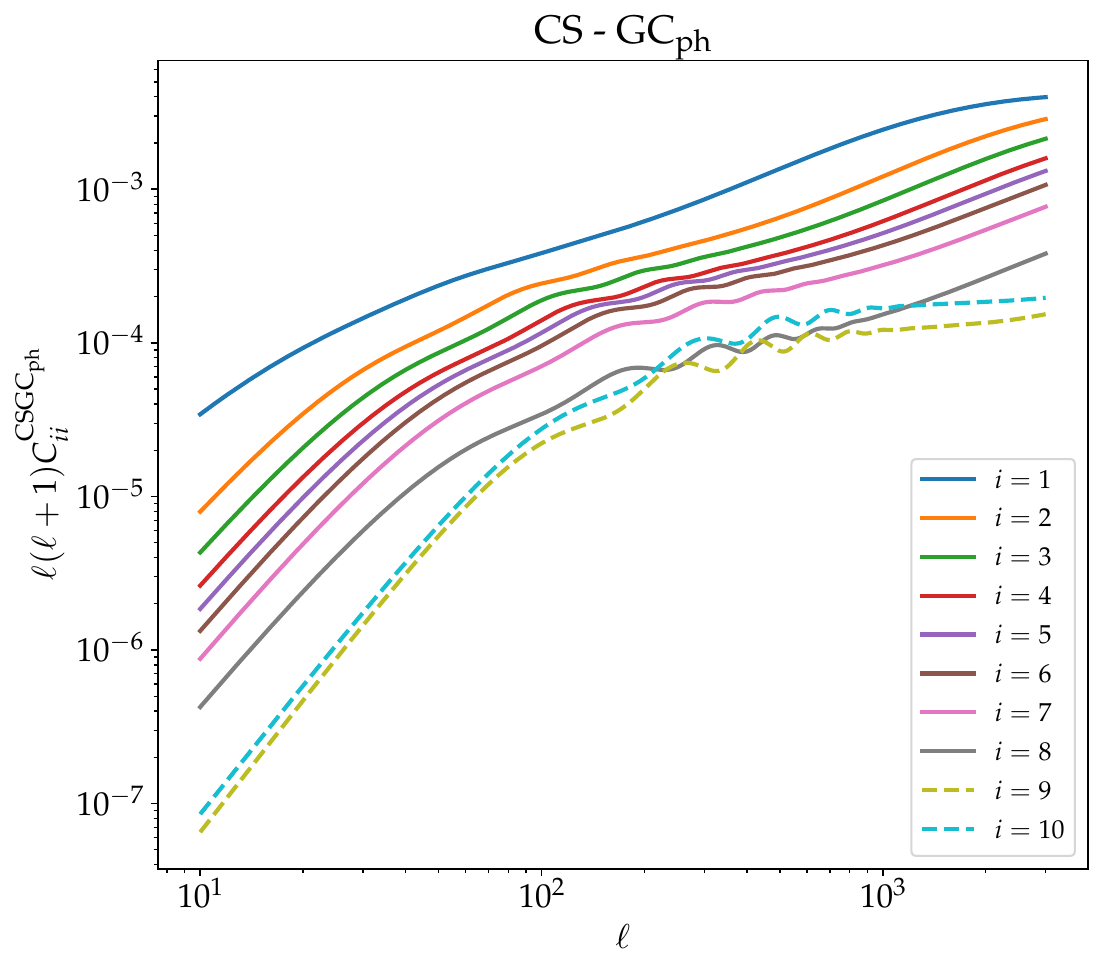}
\end{minipage}
\caption{Angular power spectra as a function of $\ell$ and for the $i$th-bin with $i=1,...,10$. From top to bottom: auto-correlated angular power spectra for CS, GC$_{\rm ph}$ and GW and cross-correlated angular power spectra for the three probes.}
\label{fig:Cls}
\end{figure}

While the use of complementary probes is well known to be essential to provide tight limits on cosmological parameters by breaking degeneracies and possibly mitigate systematics, the joint analysis of correlated probes represents a rather new field in cosmology. By cross-correlating galaxy clustering, galaxy weak lensing and GW magnification, one can exploit their complementary nature to extract additional cosmological information. The cross-correlation signal quantifies the statistical relationship between the large scale matter distribution and the spatial distribution of galaxies. It allows the extraction of cosmological information that is not easily accessible by analysing each observable separately and helps to mitigate various systematic effects that can impact each probe when analysed independently. \\
The cross-correlation angular power spectra simply combine the three different probes, as in eq.~\eqref{eqz:Cl}, with (A, B) = (CS, GC$_{\text{ph}}$, GW). In this case it can be assumed that the Poisson errors are uncorrelated, which yields $N_{ij}^{\text{AB}}=0$.\\
In the right image of the second row and in the third row of figure~\ref{fig:Cls}, we show angular cross-spectra between these three probes.

\section{Forecasts: combining galaxy surveys \& GW detectors}
\label{sec:2}
In this section we show the results of the Fisher matrix analysis performed to estimate the cosmological parameter errors from the combination of Stage-IV galaxy surveys with GW detectors. We have explored the benefits that the combination of three distinct probes can deliver in the parameter constraint for different cosmological models, which take into account the presence of massive neutrinos and a dynamical DE. Along with the corresponding parameter errors, we report the contour plots giving a visible insight of the error values. 

\subsection{Settings}
The cosmological parameters of interest and their fiducial parameters are illustrated in table~\ref{tab:fiducial}. We have performed the derivatives of the $C(\ell)$ with respect to the following parameter set:
\begin{equation*}
    \theta_{\alpha}=\left\{\Omega_m, \Omega_b, w_0, w_a, h, n_s, \sigma_8, M_{\nu}, \eta_{\text{IA}}, \beta_{\text{IA}}, \mathcal{A}_{\text{IA}}, b_i\right\}
\end{equation*}
The parameters in the above list which do not appear in table~\ref{tab:fiducial} are considered as nuisance parameters, i.e. they are marginalised over and, therefore, they contribute to the errors of the cosmological parameters.
For such an analysis we have considered four scenarios:
\begin{itemize}
    \item the $\nu\Lambda$CDM model with fixed and varying total neutrino mass, $M_{\nu}$
    \item the $\nu w_0w_a$CDM model with fixed and varying $M_{\nu}$
\end{itemize}
Hereafter, in the $\Lambda$CDM case $\Omega_{\text{DE}}\equiv\Omega_{\Lambda}$, meaning that the equation of state parameters are fixed at $w_0 =-1$ and $w_a=0$, and we did not consider the derivatives of the $C(\ell)$ with respect to them in the Fisher matrix analysis. Moreover, all the results refer to a fiducial neutrino mass of $M_{\nu}=0.06$ eV, kept fixed or not in the analysis depending on the different forecasting cases.\\
We have considered two configurations, i.e. a pessimistic ($\ell^{\rm GW}_{\text{max}}=100$) and an optimistic  ($\ell^{\rm GW}_{\text{max}}=1000$) scenario for the GW probe, while, the maximum multipole for CS in the nonlinear regime is chosen to be $\ell^{\rm CS}_{\text{max}}=1500$, and for  GC$_{\text{ph}}$ $\ell^{\rm GC_{\rm ph}}_{\text{max}}=750$. 
Consequently, the cross-correlation signals are limited to the lowest of these multipoles. Therefore, the sum over $\ell$ in the implementation of eq.~\eqref{eqz:fisher} is performed within the range $\ell_{\text{min}}$ and $\ell_{\text{max}}$ depending on the probe and scenario considered, where, for every probe and for both scenarios, we set $\ell_{\text{min}}=10$. This truncation at small multipoles eliminates very large scales, at which the Limber approximation breaks down. Moreover, for the GW optimistic case, we have analysed two scenarios with a different luminosity distance error, i.e. $\sigma_{d_L}/d_L=1\%,10\%$.  In addition, we have chosen to take a set of linearly equi-spaced data-points so that the bin width appearing in eq.~\eqref{eqz:sigma_cov} is $\Delta\ell=1$. Finally, concerning the sky coverage, in the case of CS and GC$_{\text{ph}}$ we consider a covered area of $A=15000$ deg$^2$, providing the value of $f_{\text{sky}}^{\rm g}$ for the two probes, while all currently planned GW interferometers present a full-sky coverage ($f_{\text{sky}}^{\text{gw}}=1$).\\
In table~\ref{tab:settings}, we summarise the different parameter settings in order to evaluate the Fisher matrix for the three probes.
\begin{table}
\centering
\renewcommand\arraystretch{1.1}
\caption{Summary of the parameters entering the noise modelling, the binned source distributions of eq.~\eqref{eqz:n_i_z} and the multipole settings for the probes considered. We have considered two different scenarios for the GW magnification, a pessimistic and an optimistic one with $\ell^{\rm GW}_{\rm max}=100,1000$ respectively. \\}
\label{tab:settings}
\begin{tabular}{|c|c|c|c|c|c|c|} 
    \hline
\multicolumn{7}{|c|}{Cosmic Shear} \\
    \hline
     $f_{\text{sky}}^{\rm g}$ & $\sigma_{\text{b}}$ & $\sigma_{\text{CS}}$ & $z_0$ & $\bar{n}_{\rm g}$ [arcmin]$^{-2}$ & $\ell_{\text{min}}$ & $\ell_{\text{max}}$\\
    0.3636 & 0.05 & 0.3 & $0.9/\sqrt{2}$ & 30 & 10 & 1500  \\
    \hline\hline
\multicolumn{7}{|c|}{Photometric Galaxy Clustering} \\
    \hline
    $f_{\text{sky}}^{\rm g}$ & $\sigma_{\text{b}}$ &  & $z_0$ & $\bar{n}_{\rm g}$ [arcmin]$^{-2}$ & $\ell_{\text{min}}$ & $\ell_{\text{max}}$ \\
    0.3636 & 0.05 & - & $0.9/\sqrt{2}$ & 30 & 10 & 750 \\
    \hline\hline
\multicolumn{7}{|c|}{GW magnification} \\
    \hline
    $f_{\text{sky}}^{\text{gw}}$ & $\sigma_{\text{gw}}$ & $\sigma_{d_L}/d_L (\%)$ & $z_0$ & $N_\text{gw}$ & $\ell_{\text{min}}$ & $\ell_{\text{max}}$ \\ 
    1 & 0.05 & 1 - 10 & 1.5 & 10$^6$ & 10 & 100 - 1000  \\
    \hline
\end{tabular}
\end{table}

\subsection{The \texorpdfstring{$\nu\Lambda$CDM}{} scenario}
Let us start discussing the more conservative scenario of $\ell_\textrm{GW}\leq 100$. In this case, the contribution of the GW magnification to the cosmological parameter determination is small compared to that of the other probes considered; for instance, the uncertainties on $h$ and $n_s$ are larger than the one obtained by CS alone by a factor of 50 and 150, respectively. Even when fully exploiting the full constraining power of the GW magnification, cross-correlating it with all the other probes in the so-called 6$\times$2pt statistics, we see improvement on the cosmological parameter errors smaller than 1$\%$ as compared to the standard 3$\times$2pt for Stage-IV surveys.

The results are different if we consider our second scenario, with $\ell_\textrm{GW}\leq 1000$ and $\sigma_{d_L}/d_L=10\%$. The GW-only uncertainties reduce significantly, getting closer to the CS and $\textrm{GC}_\textrm{ph}$-only measurements. For instance, the GW-only error on $\Omega_b$ and $h$ decreases by a factor of 10, while that on $n_s$ by a factor of 16. When comparing the 3$\times$2-pt with the full 6$\times$2pt, the two most remarkable improvements come from $\sigma_8$ and $\Omega_m$, whose uncertainty get reduced by a factor 4 and 5$\%$, respectively. This should come as no surprise, as we are using GWs to trace the effect of lensing which is known to be most sensitive to these two parameters. In fact, the two parameters with smaller improvement are $\Omega_b$ and $h$, with an uncertainty reduction below $1\%$; also this can be explained with the same reasoning, as these are two parameters with a small effect on lensing.

Finally, we consider the $\sigma_{d_L}/d_L=1\%$ scenario. In this case the GW-only uncertainties get more and more competitive and are in fact similar to the CS and $\textrm{GC}_\textrm{ph}$-only marginalised errors. The results are even better when we consider the full 6$\times$2pt, as we obtain improvements up to 16\%, 12\% and 14\% for $\Omega_m$, $n_s$ and $\sigma_8$ respectively. All the uncertainties are reported in the first part of table~\ref{tab:errori_lcdm}.

\begin{table}
\small
\centering
\renewcommand\arraystretch{1.0}
\caption{marginalised $1-\sigma$ errors for the galaxy CS, GC$_{\text{ph}}$ and GW magnification, in the case of a cosmological constant. In the GW and 6$\times$2pt block, the first row shows the errors for the pessimistic scenario and the second and third for the optimistic one with $\sigma_{d_L}/d_L=10\%$ and $1\%$, respectively. The last row in each model block ($\Delta\%$) represents the percentage difference between the $1-\sigma$ errors from the Stage-IV $3\times2$pt and the $6\times2$pt statistics.\\}
\label{tab:errori_lcdm}
\begin{tabular}{l|c|c|c|c|c|c|c|c}
  \toprule                       
  Probe & $\Omega_m$ & $\Omega_b$ & $h$ & $n_s$ & $\sigma_8$ & $M_{\nu}$[eV] & $w_0$ & $w_a$ \\
  \midrule 
  \multicolumn{8}{c}{$\nu\Lambda$CDM (fixed $M_\nu$)} \\
\midrule
CS & 0.00558 & 0.0237 & 0.137 & 0.0321 & 0.00687 & - &  - & - \\
\midrule
GC$_{\text{ph}}$ & 0.00751 & 0.00285 & 0.0265 & 0.0292 & 0.0120 & - & - & - \\ 
\midrule
\multirow{3}{5em}{GW} & 0.0599 & 0.504 & 6.74 & 4.78 & 0.564 & - & - & - \\
                      & 0.0195 & 0.0575 & 0.402 & 0.146 & 0.0242 & - & - & - \\ 
                      & 0.00767 & 0.0257 & 0.165 & 0.0520 & 0.00936 & - & - & - \\
\midrule
3$\times$2pt$_{\rm Stage-IV}$ & 0.00256 & 0.00242 & 0.0172 & 0.00810 & 0.00311 & - & - & - \\ 
\midrule
\multirow{3}{5em}{6$\times$2pt} & 0.00253 & 0.00242 & 0.0172 & 0.00804 & 0.00308 & - & - & - \\ 
                            & 0.00245 & 0.00241 & 0.0171 & 0.00782 & 0.00300 & - & - & - \\ 
                            & 0.00214 & 0.00237 & 0.0166 & 0.00710 & 0.00268 & - & - & - \\
\midrule
\multirow{3}{5em}{$\Delta$\%} & 0.95 & 0.13 & 0.10 & 0.67 & 0.79 & - & - & - \\ 
                            & 4.26 & 0.45 & 0.88 & 3.47 & 3.54 & - & - & - \\
                            & 16.3 & 1.95 & 3.56 & 12.3 & 13.8 & - & - & - \\

\midrule
\multicolumn{8}{c}{$\nu\Lambda$CDM}\\
\midrule
CS & 0.00610 & 0.0253 & 0.140 & 0.0334 & 0.0134 & 0.272 & - & - \\
\midrule
GC$_{\text{ph}}$ & 0.00769 & 0.00326 & 0.0330 & 0.0416 & 0.0244 & 0.126 & - & -\\
\midrule
\multirow{3}{5em}{GW} & 0.0630 & 0.585 & 7.86 & 5.92 & 0.717 & 2.86 & - & - \\
                      & 0.0203 & 0.0623 & 0.402 & 0.150 & 0.0392 & 0.642 & - & -\\
                      & 0.00801 & 0.0275 & 0.165 & 0.0525 & 0.0161 & 0.297 & - & - \\

\midrule
3$\times$2pt$_{\rm Stage-IV}$ & 0.00305 & 0.00242 & 0.0176 & 0.00810 & 0.00511 & 0.0662 & - & - \\ 

\midrule

\multirow{3}{5em}{6$\times$2pt} & 0.00301 & 0.00242 & 0.0175 & 0.00804 & 0.00507 & 0.0657 & - & - \\ 
                            & 0.00289 & 0.00241 & 0.0174 & 0.00782 & 0.00489 & 0.0647 & - & - \\
                            & 0.00245 & 0.0237 & 0.0170 & 0.00710 & 0.00426 & 0.0594 & - & - \\
\midrule
\multirow{3}{5em}{$\Delta$\%} & 1.02 & 0.13 & 0.15 & 0.66 & 0.77 & 0.69 & - & - \\ 
                            & 5.22 & 0.45 & 0.73 & 3.47 & 4.25 & 2.21 & - & - \\
                            & 19.7 & 1.97 & 2.95 & 12.3 & 16.7 & 10.2 & - & - \\
  \bottomrule                    
\end{tabular}
\end{table}

\subsection{The \texorpdfstring{$\nu\Lambda$CDM}{} scenario with free \texorpdfstring{$M_{\nu}$}{}}
Here we consider our first extension to the standard cosmological model with $\Lambda$ and minimum total neutrino mass $M_\nu=0.06$ eV, namely when we add the neutrino mass $M_\nu$ as a free parameter. As in the previous section, we start considering the pessimistic case. Except for $\sigma_8$, the addition of the total neutrino mass, $M_{\nu}$, in the Fisher matrix analysis as a free parameter, slightly increases all the marginalised errors, which in this case are reported in the second part of table~\ref{tab:errori_lcdm}. Indeed, the $\sigma_8$ constraint is weakened by a factor of $\sim$ 2  with respect to the $M_\nu$-fixed case. This is actually expected since the effects of $\sigma_8$ and massive neutrinos on the matter power spectrum are similar, i.e. they are degenerate parameters: $\sigma_8$ controls the normalisation of $P_{mm}$, and $M_{\nu}$ suppresses the $P_{mm}$ amplitude in a scale-dependent way due to neutrinos free streaming~\cite{LESGOURGUES_2006}. The second most affected parameter, when at one between CS and GW is considered, is $\Omega_m$; this is an indirect effect, due to the correlation of the matter density with $\sigma_8$. Also in this case the GW-only constraints are way weaker than the ones obtained considering the other two probes and the main improvement given by the 6$\times$2pt over the 3$\times$2pt is of $1\%$ for $\sigma_8$.

In a similar fashion to the previous case, when we consider $\sigma_{d_L}/d_L=10\%$, the GW-only constraints get better and the 6$\times$2pt statistics offer some improvements over the standard 3$\times$2pt ones. Finally, when we set $\sigma_{d_L}/d_L=1\%$, the GW-only constraints get competitive with the other single probes constraints and the 6$\times$2pt offer a nice improvement over the standard analysis, improving the constraints over $\sigma_8$ and $\Omega_m$, respectively, by 16 and $20\%$, while the error on $M_\nu$ is decreased by $10\%$.

\begin{table}[hbt!]
\small
\centering
\renewcommand\arraystretch{1.0}
\caption{marginalised $1-\sigma$ errors for the galaxy CS, GC$_{\rm ph}$ and magnification of GW probes for the dynamical DE scenarios. In the GW and 6$\times$2pt block, the first row shows the errors for the pessimistic scenario and the second and third for the optimistic one with $\sigma_{d_L}/d_L=10\%$ and $1\%$ respectively. The last row in each model block ($\Delta\%$), represents the percentage difference between the $1-\sigma$ errors from the Stage-IV $3\times2$pt and the $6\times2$pt statistics.\\}
\label{tab:errori_w0wa}
\begin{tabular}{l|c|c|c|c|c|c|c|c}
  \toprule                      
  Probe & $\Omega_m$ & $\Omega_b$ & $h$ & $n_s$ & $\sigma_8$ & $M_{\nu}$[eV] & $w_0$ & $w_a$ \\
  \midrule 
\multicolumn{8}{c}{$\nu w_0w_a$CDM (fixed $M_\nu$)}\\
\midrule
CS & 0.0140 & 0.0238  & 0.138 & 0.0348 & 0.0155 & - & 0.163 & 0.592\\ 
                    
\midrule
GC$_{\text{ph}}$ & 0.0482 & 0.00942 & 0.0491 & 0.0432 & 0.0429 & - & 0.569 & 1.83 \\ 
                                
\midrule
\multirow{3}{5em}{GW} & 0.650 & 0.624 & 7.54 & 5.73 & 1.34 & - & 5.63 & 21.4 \\ 
                      & 0.161 & 0.0619  & 0.548 & 0.256 & 0.167 & - & 1.45 & 5.14 \\
                      & 0.0551 & 0.0276 & 0.197 & 0.0830 & 0.0579 & - & 0.503 & 1.81 \\
\midrule
3$\times$2pt$_{\rm Stage-IV}$ & 0.00349 & 0.00259 & 0.0197 & 0.0100 & 0.00403 & - & 0.0420 & 0.172 \\ 

\midrule

\multirow{3}{5em}{6$\times$2pt} & 0.00346 & 0.00258 & 0.0197 & 0.00992 & 0.00401 & - & 0.0418 & 0.172 \\ 
                            & 0.00336 & 0.00256 & 0.0196 & 0.00968 & 0.00395 & - & 0.0400 & 0.165 \\
                            & 0.00304 & 0.00250 & 0.0191 & 0.00889 & 0.00372 & - & 0.0344 & 0.145 \\
\midrule
\multirow{3}{5em}{$\Delta$\%} & 0.70 & 0.23 & 0.14 & 0.80 & 0.42 & - & 0.57 & 0.50 \\ 
                            & 3.64 & 0.85 & 0.73 & 3.18 & 2.03 & - & 4.87 & 4.42 \\
                            & 12.9 & 3.32 & 2.90 & 11.0 & 7.64 & - & 18.0 & 16.1 \\
\midrule
\multicolumn{8}{c}{$\nu w_0w_a$CDM}\\
\midrule
CS & 0.0155 & 0.0258 & 0.144 & 0.0413 & 0.0266 & 0.350 & 0.164 & 0.652 \\ 

\midrule
GC$_{\text{ph}}$ & 0.0486 & 0.00945 & 0.0561 & 0.0545 & 0.0513 & 0.128 & 0.570 & 1.83 \\ 
                                
\midrule
\multirow{3}{5em}{GW} & 0.790 & 0.631 & 7.92 & 6.05 & 1.39 & 3.82 & 6.65 & 26.5 \\ 
                      & 0.209 & 0.0666 & 0.695 & 0.315 & 0.251 & 0.966 & 1.79 & 7.01 \\
                      & 0.0705 & 0.0288 & 0.233 & 0.0996 & 0.0876 & 0.429 & 0.612 & 2.41 \\
\midrule
3$\times$2pt$_{\rm Stage-IV}$ & 0.00453 & 0.00269 & 0.0204 & 0.0118 & 0.00843 & 0.0967 & 0.0428 & 0.200 \\ 
            
\midrule

\multirow{3}{5em}{6$\times$2pt} & 0.00449 & 0.00268 & 0.0204 & 0.0117 & 0.00836 & 0.0956 & 0.0426 & 0.198 \\ 
                            & 0.00440 & 0.00267 & 0.0202 & 0.0115 & 0.00828 & 0.0949 & 0.0408 & 0.193 \\
                            & 0.00404 & 0.00259 & 0.0197 & 0.0106 & 0.00771 & 0.0872 & 0.0355 & 0.173 \\
\midrule
\multirow{3}{5em}{$\Delta$\%} & 0.74 & 0.22 & 0.17 & 0.73 & 0.74 & 1.08 & 0.62 & 0.71 \\ 
                            & 2.71 & 0.92 & 0.84 & 2.90 & 1.76 & 1.88 & 4.65 & 3.58 \\
                            & 10.8 & 3.53 & 3.40 & 10.7 & 8.46 & 9.76 & 17.2 & 13.3 \\

  \bottomrule                  
\end{tabular}
\end{table}

\subsection{The \texorpdfstring{$\nu w_0w_a$CDM}{} scenario}
In this section we consider a second extension for the standard cosmological model, when we consider as additional free parameters $w_0$ and $w_a$ in the DE equation of state. We keep the neutrino mass fixed, an assumption that will be relaxed in the next section. Results on the marginalised errors are shown in the first part of table~\ref{tab:errori_w0wa}.

One can observe that varying $w_0$ and $w_a$ increases all errors but it mainly affects $\Omega_m$ and $\sigma_8$. The impact on $\Omega_m$ can be explained considering that $w_0$ and $w_a$ enter the Hubble parameter $E(z)$ together with $\Omega_m$. Additionally, $w_0$ and $w_a$ affect the linear growth factor and therefore can affect the normalisation of the matter power spectrum, an effect degenerate with $\sigma_8$. Also in this case the GW-magnification contribution is only marginal, both when considering the single probes uncertainties and when comparing the 6x2-pt with the standard 3$\times$2-pt statistics.

When we move to the more optimistic settings of $\ell_\textrm{max}=1000$ with $\sigma_{d_L}/d_L=10\%$, we see as before a general improvement, slightly reduced by the introduction of the varying DE equation of state parameters. Finally, when we consider the $\sigma_{d_L}/d_L=1\%$ case, the GW contribution helps in determining the cosmological parameters, with most of the uncertainties being reduced by $10\%$ or more using the full $6\times 2$-pt. In particular, $w_0$ and $w_a$ marginalised errors improve respectively by $17$ and $13\%$; this shows the importance of additional cross-correlations, especially when opening the parameter space as this can introduce parameter degeneracies that can be alleviated with additional probes.

\subsection{The \texorpdfstring{$\nu w_0w_a$CDM}{} scenario with free \texorpdfstring{$M_{\nu}$}{}}
\begin{figure}[H]
\centering
\includegraphics[width=0.9\textwidth]{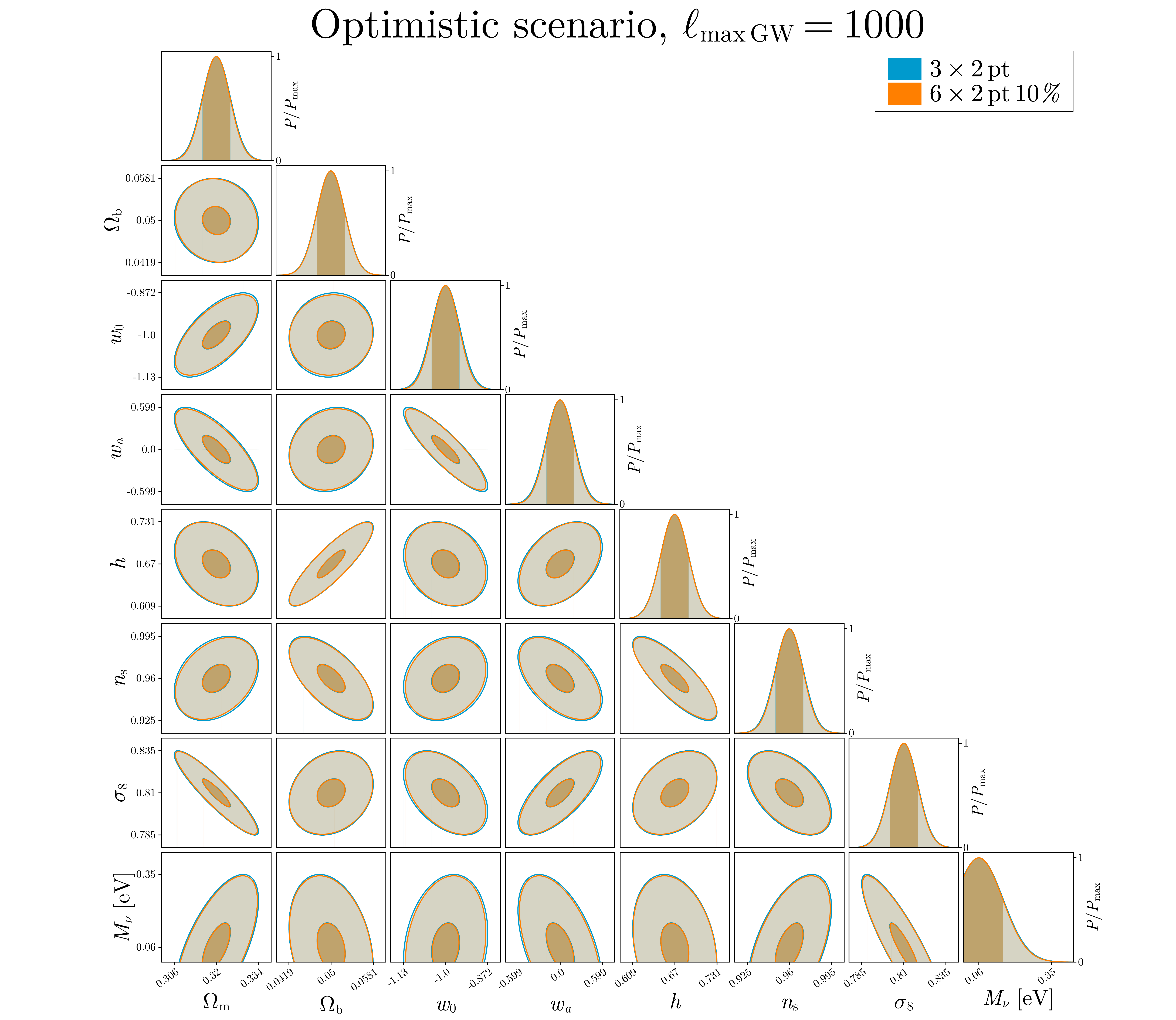}
\includegraphics[width=0.9\textwidth]{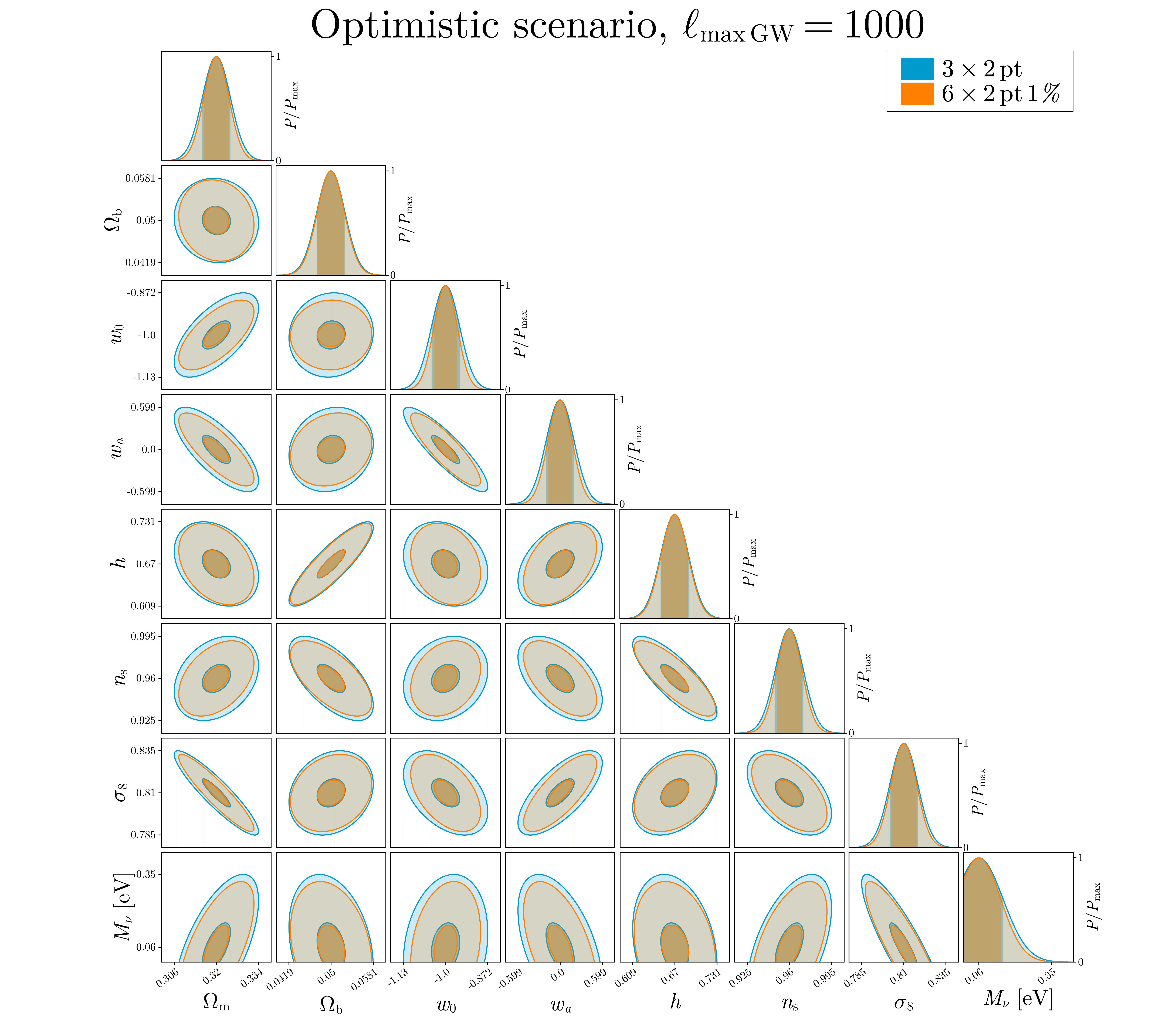}
\caption{Fisher matrix marginalised contours for the $\nu w_0w_a$CDM model with varying $M_{\nu}$, in the optimistic scenario with different luminosity distance error, i.e. $d_L=10\%$ (top) and $d_L=1\%$ (bottom).}
\label{fig:contour_6x2_nuw0wa_optim}

\end{figure}

\begin{figure}[H]
\centering
\includegraphics[width=0.95\textwidth]{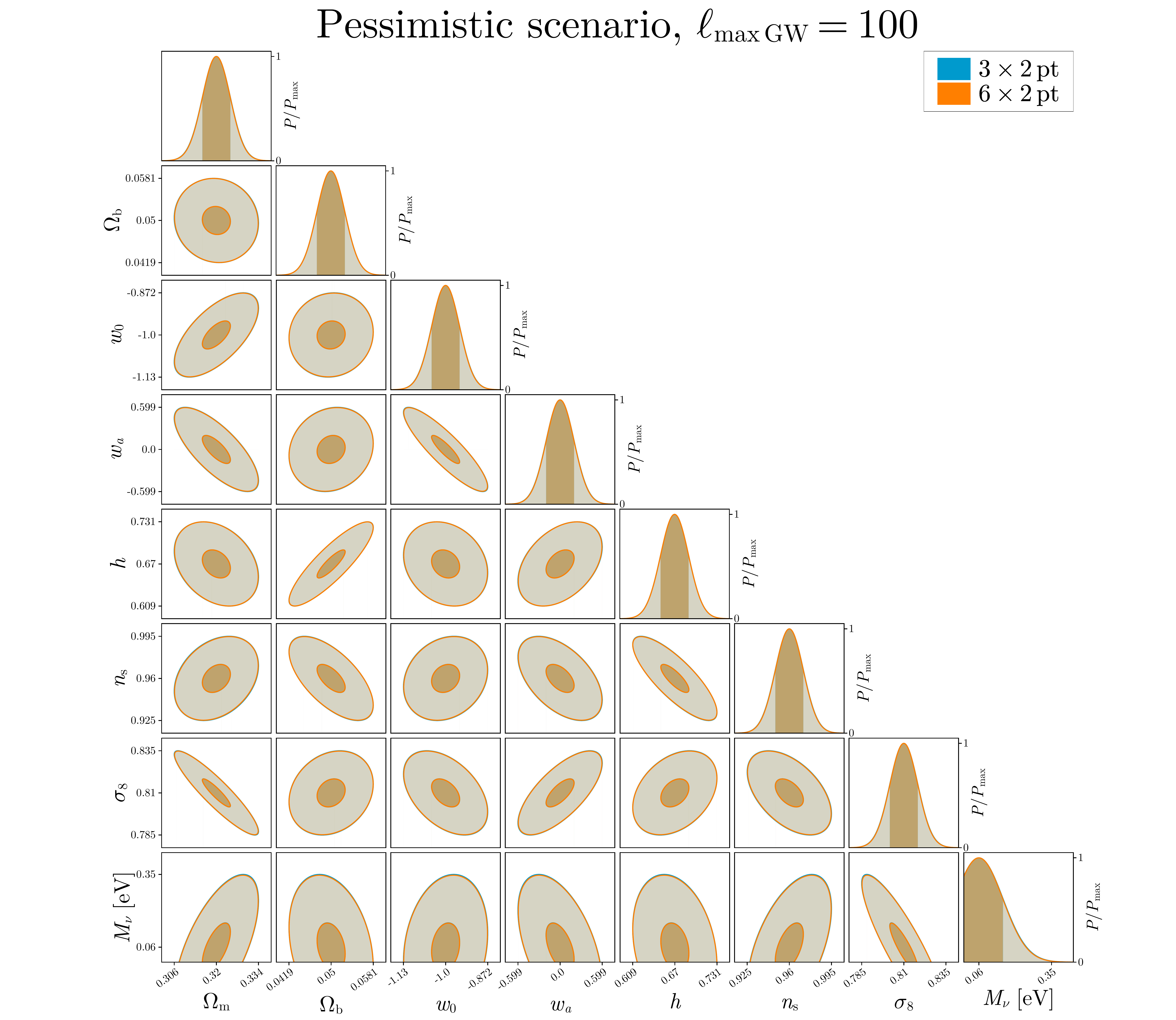}
\caption{Fisher matrix marginalised contours for the $\nu w_0w_a$CDM model with varying $M_{\nu}$, in the pessimistic scenario.}
\label{fig:contour_6x2_nuw0wa_pessim}
\end{figure}

Finally, here we present the results when we consider our most general extension of standard cosmological model, namely the $\nu w_0 w_a$CDM scenario, when we consider as free parameters both the neutrino mass and the DE equation of state parameters. The values of the cosmological parameter constraints from the various probes considered are reported in the second part of table~\ref{tab:errori_w0wa}, while the contour plots for this model cosmology are shown in figure~\ref{fig:contour_6x2_nuw0wa_pessim} for the pessimistic case and in figure~\ref{fig:contour_6x2_nuw0wa_optim}, for the optimistic case with different value of $\sigma_{d_L}/d_L$. 

\begin{figure}[ht]
\centering
\includegraphics[width=0.52\textwidth]{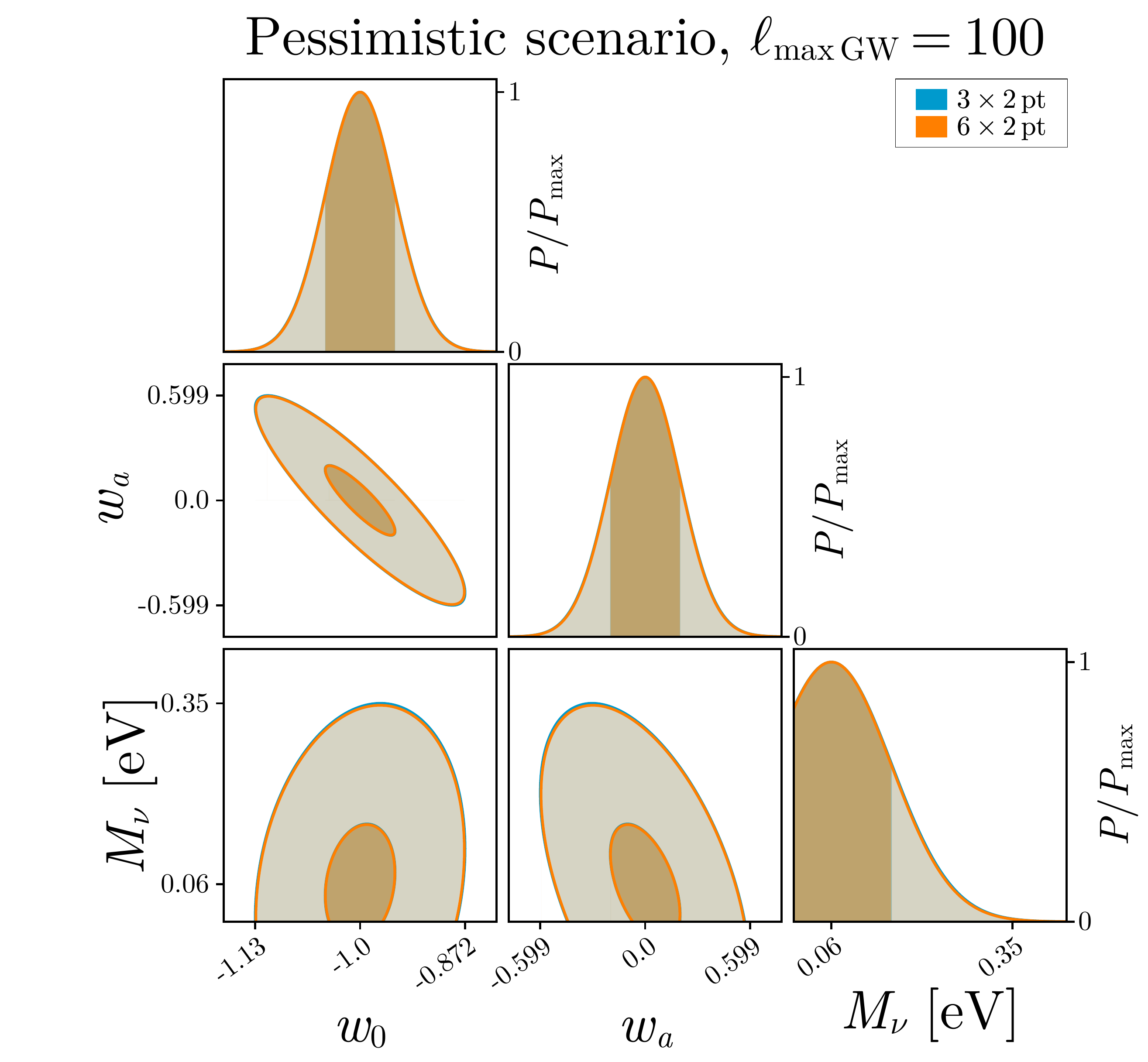}
\setlength{\tabcolsep}{0.01pt}
\begin{tabular}{cc}
    \includegraphics[width=0.52\textwidth]{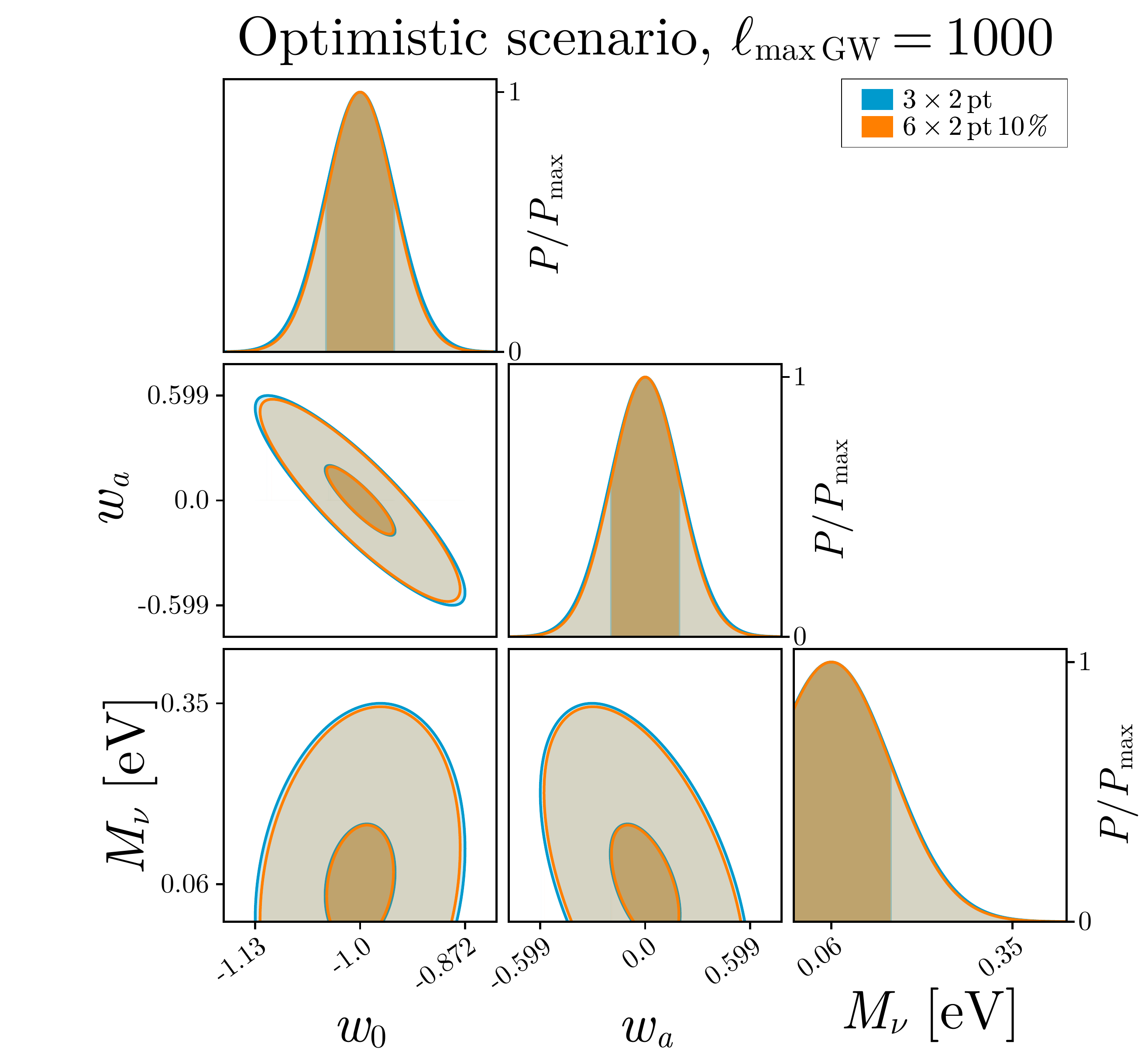} &
    \includegraphics[width=0.52\textwidth]{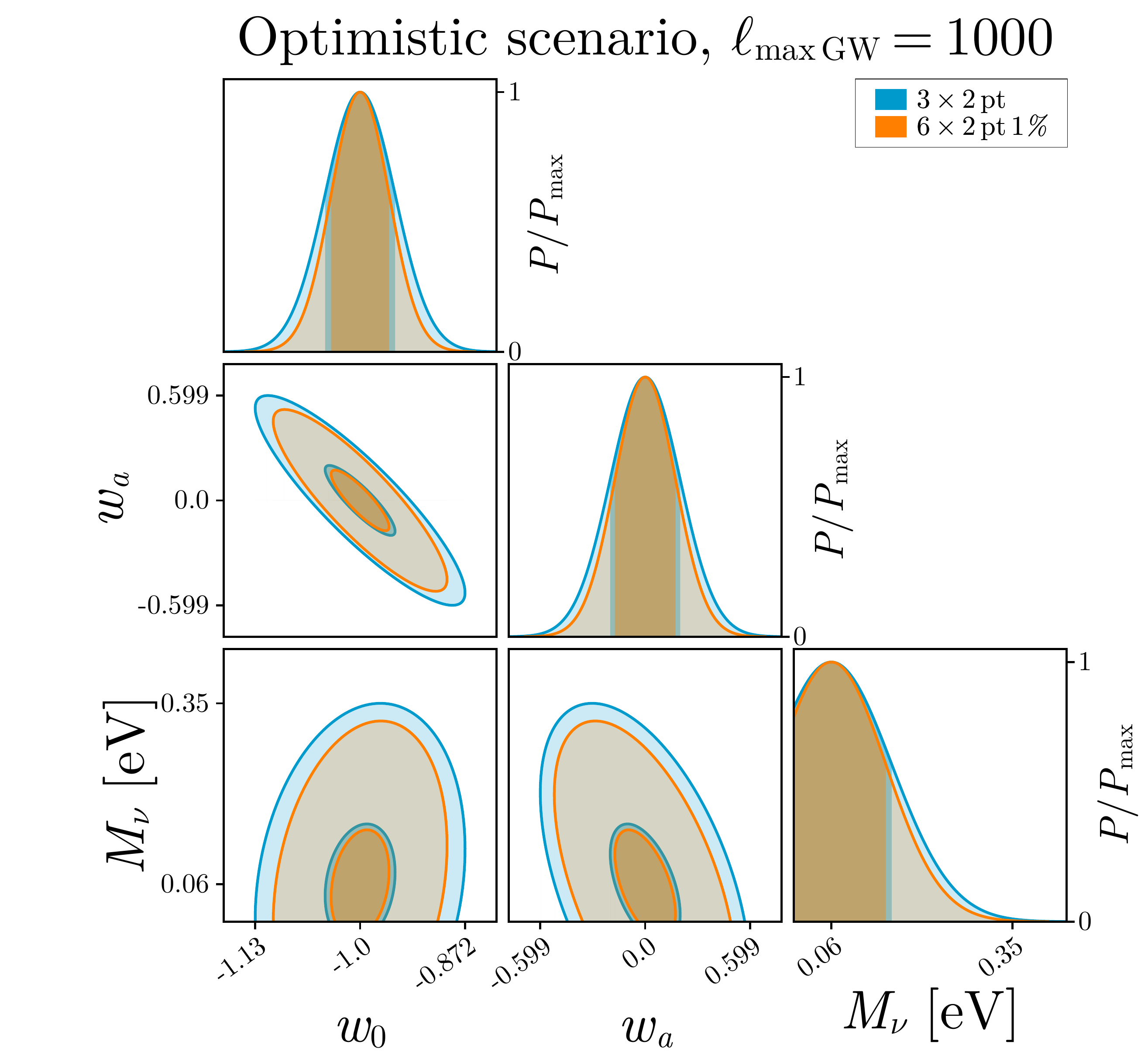}
\end{tabular}
\caption{Fisher matrix marginalised contours for the $\nu w_0w_a$CDM model with varying $M_{\nu}$, in the pessimistic (top) and optimistic (bottom) scenarios with $\ell^{\rm GW}_{\text{max}}=100$ and $\ell^{\rm GW}_{\text{max}}=1000$, respectively. We have also considered a different luminosity distance error for the GW probe for the optimistic scenarios i.e. $\sigma_{d_L}/d_L=10\%$ (left) and $\sigma_{d_L}/d_L=1\%$ (right).}
\label{fig:contour_6x2_tr_nuw0wa}
\end{figure}

As for the previous scenarios considered, the contribution of the GW magnification in the pessimistic case is only marginal. Also, we can see the effect of adding $M_\nu$ as a free parameter, which affects most of the uncertainties shown in table~\ref{tab:errori_w0wa}, which we might have expected as we are introducing additional degeneracies among the free parameters. When moving to the more optimistic setting of $\ell^{\rm GW}_\textrm{max}=1000$ and $\sigma_{d_L}/d_L=10\%$, we find that the $6\times2$pt statistics give some contributions to the $3\times2$pt ones, but smaller than in the previous cases. Finally, when we move to our most optimistic scenario of $\sigma_{d_L}/d_L=1\%$, the $6\times2$pt statistics give a sizeable contribution, with most of the parameters getting an improvement of $\sim 10\%$. Concretely, $M_\nu$, $w_0$, and $w_a$ get, respectively, an improvement of 10, 17, and $13\%$, showing also in this scenario the ability of additional probes to break degeneracies among parameters. Figure~\ref{fig:contour_6x2_tr_nuw0wa} shows a closer focus on the DE equation of state parameters and the total neutrino mass in both pessimistic and optimistic scenarios.

A notable fact is that the errors on the cosmological parameters decrease with the progressive increasing of $\ell^{\rm GW}_{\rm max}$, as shown in figure~\ref{fig:err_cum}. However the shot noise of the GW magnification has a larger influence on the signal with respect to the other probes as figure~\ref{fig:noise} shows. In fact, going up to $\ell^{\rm GW}_{\text{max}}>1000$ would result in a further enhancement of the signal-to-noise ratio. This on one hand would imply tighter constraints on cosmological parameters when all sources are combined. At the same time though, the GC$_{\text{ph}}$ and CS signal themselves would benefit from the accessibility of higher multipoles, resulting in tighter galaxy-only constraints to begin with. 
Thus, we expect that qualitatively, the impact of GW on the cosmological parameters remains similar even when larger multipoles are accessible. 

\section*{Conclusions}\label{sec:clc}
\addcontentsline{toc}{section}{Conclusions}
\markboth{Conclusions}{}
In this work we have investigated the synergies between Stage-IV galaxy surveys and future GW observatories for constraining the underlying cosmological model of the Universe, focussing on photometric galaxy clustering, cosmic shear and GW magnification as cosmological probes. By tracing the underlying matter field, the latter represents a probe able to contribute to cosmological parameter constraints in a complementary way with respect to galaxy surveys, and possibly also to minimise systematics effects via their cross-correlation.

\noindent In order to evaluate the Fisher matrix described in eq.~\eqref{eqz:fisher}, we have implemented a code that computes the angular power spectra, $C(\ell)$, and their derivatives with respect to cosmological and nuisance parameters, for each of the probes considered.
Before the evaluation of the full $6\times2$pt statistics composed by the combination and cross-correlation of the different $C(\ell)$, we have validated our code for the $3\times2$pt statistics against \cite{Euclid_2020}, as shown in appendix~\ref{apndx_A}.

\noindent After this validation, we have proceeded in section~\ref{sec:1} with the evaluation of the $6\times2$pt statistics given by the combination of Stage-IV galaxy survey with GW magnification from future detectors, and with the implementation of the associated Fisher matrix. In section~\ref{sec:2} we present parameter forecasts for the different cosmological models considered, starting from the standard cosmological $\nu\Lambda$CDM model with fixed $M_\nu=0.06$. We have then let $M_\nu$ vary as a free parameter. 

As a further step, we have considered the case of a DE model with varying equation of state, as opposite to the cosmological constant $\Lambda$ case. In particular, following the so-called CPL parameterisation, we have included $w_0$ and $w_a$ to the parameter space of the Fisher matrix analysis. Then, we let $M_\nu$, $w_0$ and $w_a$ vary simultaneously as free parameters.

\noindent The main results of this work are reported in tables.~\ref{tab:errori_lcdm} and~\ref{tab:errori_w0wa} of section~\ref{sec:2}. 
We have considered two configurations, i.e. a pessimistic ($\ell^{\rm GW}_{\text{max}}=100$) and an optimistic  ($\ell^{\rm GW}_{\text{max}}=1000$) scenario for the GW probe, while for CS in the nonlinear regime we have chosen $\ell^{\rm CS}_{\text{max}}=1500$, and $\ell^{\rm GC_{\rm ph}}_{\text{max}}=750$ for  GC$_{\text{ph}}$.

In addition, for the optimistic scenario we have considered two different values of the luminosity distance error, i.e. $\sigma_{d_L}/d_L=1\%\,,10\%$, as expected from forthcoming GW observatories. Our goal was to understand if/when the GW magnification can provide a substantial contribution to the constraints on cosmological parameters. We have found that such a contribution is very weak, almost below 1\%, in the pessimistic case.
However, if we consider the optimistic scenario with $\ell^{\rm GW}_{\rm max}=1000$, the GW magnification provides a larger constraining power, especially for the DE equation of state parameters and the total neutrino mass. In fact, when $\sigma_{d_L}/d_L=1\%$, by combining the GW magnification and the $3\times2$pt statistics from Stage-IV surveys, all the parameter errors are reduced by $3\%$-$5\%$, and when $\sigma_{d_L}/d_L=1\%$ the improvement increases up to $10\%$-$18\%$ for $M_\nu$, $w_0$ and $w_a$.

Concluding, we have found that, in the case of future gravitational wave observatories reaching $\ell^{\rm GW}_{\rm max}=1000$, as e.g. described in~\cite{baker2019high}, the inclusion of the gravitational wave magnification can improve Stage-IV galaxy surveys performance on constraining the underlying cosmological model of the Universe.

However, here we have performed an unavoidably optimistic Fisher matrix analysis. For a final answer, a much more detailed and realistic approach will be needed, especially by including systematic effects.


\bibliography{biblio}
\bibliographystyle{JHEP}

\appendix
\section{The Fisher matrix approach}
\label{apndx_A}
\begin{figure}[H]
\centering
	\includegraphics[width=0.89\textwidth]{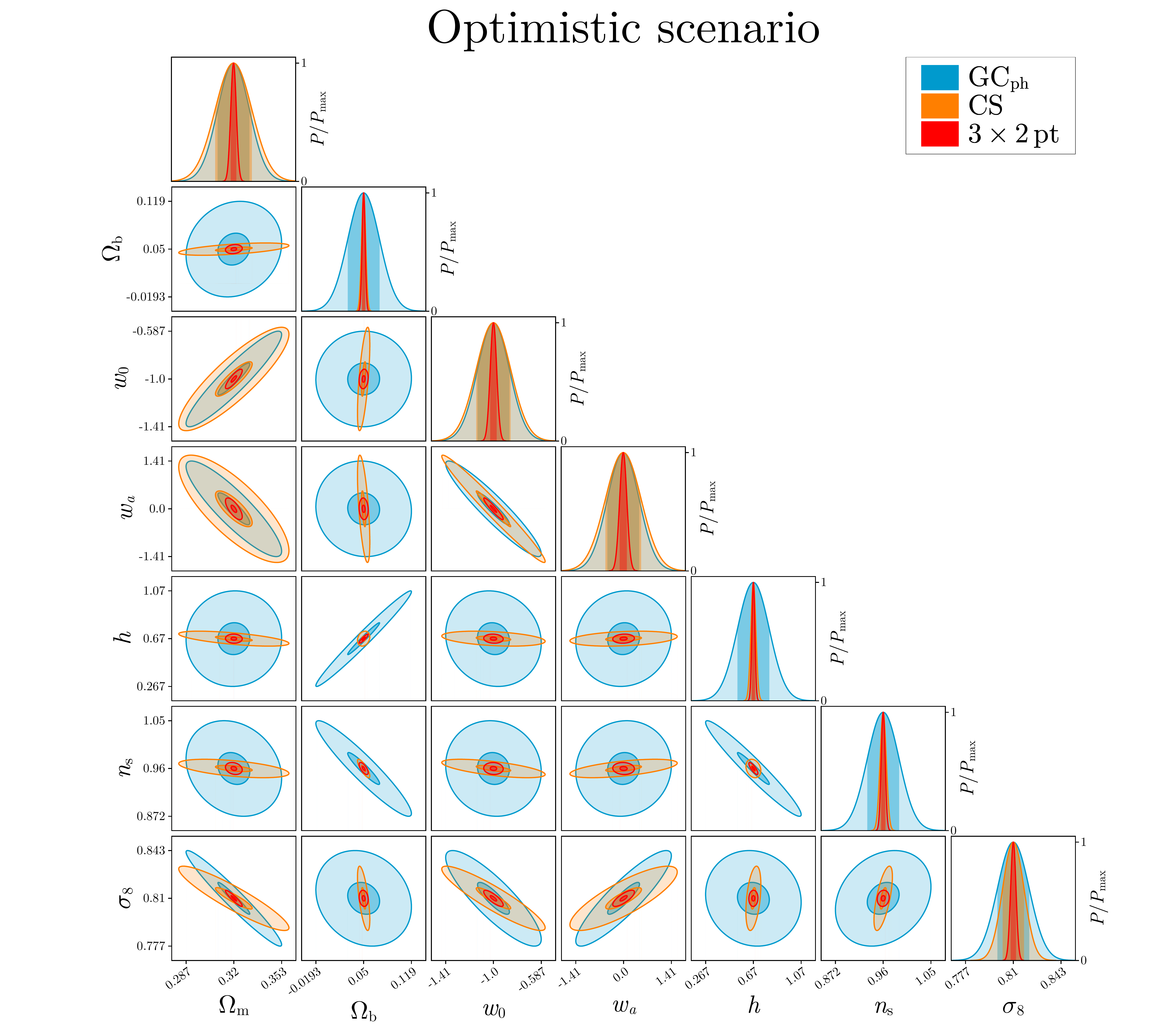} 
	\includegraphics[width=0.89\textwidth]{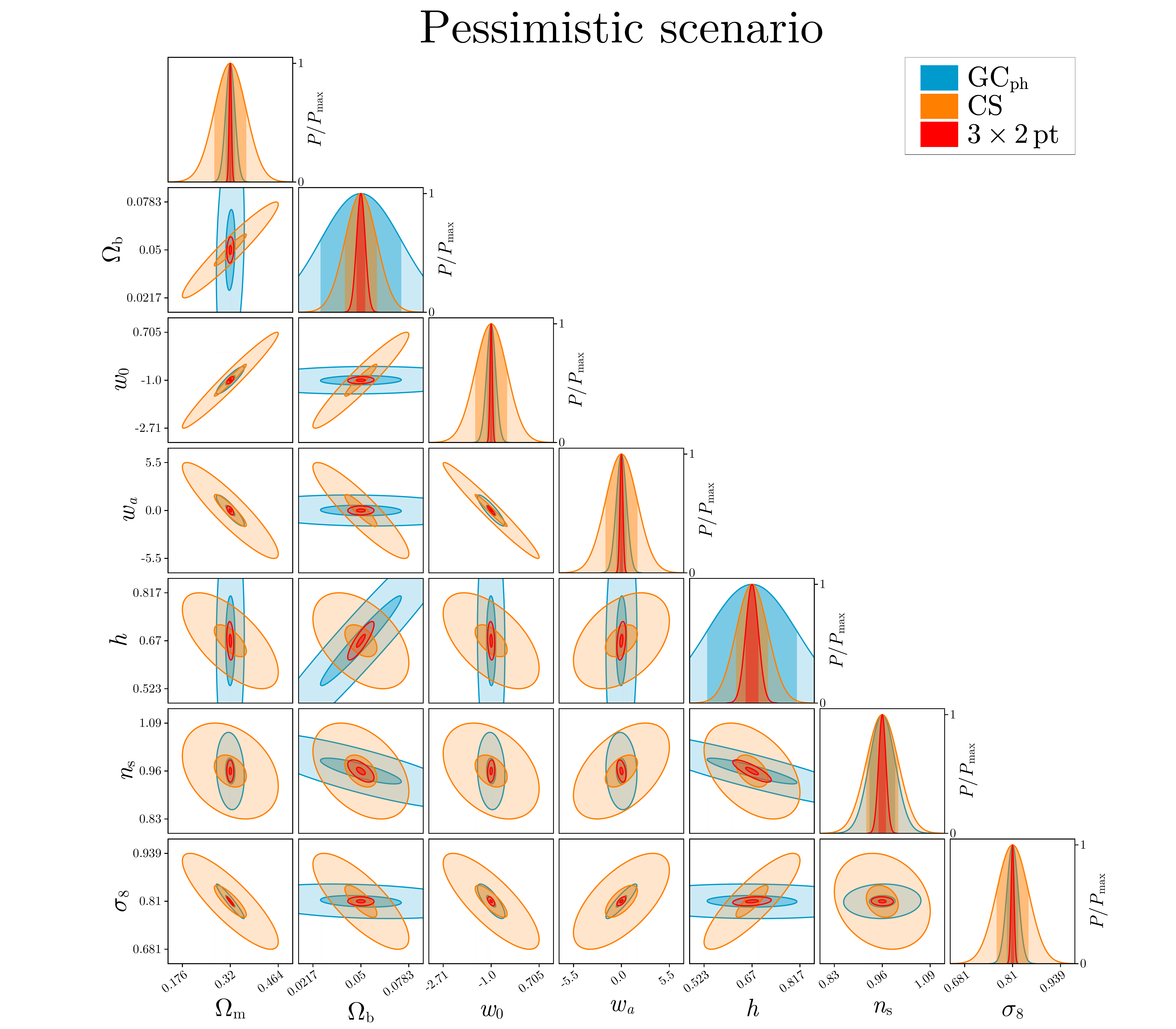}
\caption{Stage-IV forecast contours for the $\nu w_0w_a$CDM model with fixed $M_{\nu}$ in the optimistic (top) and in the pessimistic (bottom) scenario.}
\label{fig:contour_3x2}
\end{figure}

The Fisher matrix~\cite{fisher} is a mathematical tool used in statistics and parameter estimation to quantify the expected uncertainty in the measurement of a set of parameters. 
Given a data vector $\mathbf{x}$, one can look for the posterior distribution of the vector of parameters $\boldsymbol{\theta}$. By using Bayes' theorem, the posterior can be obtained as
\begin{equation}
    P(\boldsymbol{\theta}|\mathbf{x})=\frac{L(\mathbf{x}|\boldsymbol{\theta})P(\mathbf{\theta})}{P(\mathbf{x})}
\end{equation}
where $P(\boldsymbol{\theta})$ is the prior information on our parameters, $P(\mathbf{x})$ is the evidence, and $L(\mathbf{x}|\boldsymbol{\theta})$ is the likelihood of the data vector given the parameters.
The Fisher matrix is defined as the expectation value of the second derivative of the logarithm of the likelihood function with respect to the parameters of interest:
\begin{equation}
    F_{\alpha\beta}=-\left<\frac{\partial^2\ln L}{\partial\theta_{\alpha}\partial\theta_{\beta}}\right>
\end{equation}
where $\alpha$ and $\beta$ label the parameters of interest $\theta_{\alpha}$ and $\theta_{\beta}$.
It provides a way to estimate the covariance matrix of the parameter uncertainties, which describes the expected correlations between the different parameters. The inverse of the Fisher matrix, known as the covariance matrix, can be used to estimate the expected error bars on each parameter:
\begin{equation}
    C_{\alpha\beta}=\left(F\right)^{-1}_{\alpha\beta} \qquad \text{and} \qquad \sigma_{\alpha\alpha}=\sqrt{C_{\alpha\alpha}}.
\end{equation}
The square root of the reciprocal of the appropriate diagonal element of the Fisher matrix i.e. $\sigma_{\alpha}=1/F_{\alpha\alpha}$ are the unmarginalised expected errors. It can also be defined the correlation coefficient between the errors on our cosmological parameters as $\rho$, which contributes to the off-diagonal elements in the parameter covariance matrix,
\begin{equation}
    C_{\alpha\beta}=\rho_{\alpha\beta}\sigma_{\alpha}\sigma_{\beta}.
\end{equation}
If $\alpha$ and $\beta$ are completely independent, $\rho_{\alpha\beta}=0$. In order to ‘marginalise out’ a subset of the parameters one should simply remove the rows and columns in the full parameter covariance matrix that correspond to the parameters one would like to marginalise over. As we deal with angular power spectra, we can introduce the matrix $\Sigma^{\text{AB}}_{ij}(\ell)$ associated with a given $\tilde{C}(\ell)$:
\begin{equation}
\label{eqz:sigma_cov}
    \Sigma^{\text{AB}}_{ij}(\ell)=\sqrt{\frac{2}{(2\ell+1)\Delta\ell f^a_{\text{sky}}}}\left(C_{ij}^{\text{AB}}(\ell)+N_{ij}^{\text{AB}}\right),
\end{equation}
where $\Delta\ell$ is the multipole bin width, and $f^a_{\text{sky}}$ the sky fraction covered by the survey, with $a=$(g, gw). If we are dealing with a single probe (A$=$B), the covariance matrix of the $a_{\ell m}$ is given by eq.~\eqref{eqz:sigma_cov}. If two or more probes are combined together, like in the cross-correlation, all probes must be combined as
\begin{equation}
    \boldsymbol{\Sigma}(\ell) = 
    \begin{pmatrix}
    \Sigma^{\text{A}_0\text{A}_0}(\ell) & \dots & \Sigma^{\text{A}_0\text{A}_n}(\ell)\\
    \vdots & \ddots \\
    \Sigma^{\text{A}_n\text{A}_0}(\ell) & \dots & \Sigma^{\text{A}_n\text{A}_n}(\ell)
\end{pmatrix}.
\end{equation}
The same structure is shared by the $C_{\ell}$ matrix yielding:
\begin{equation}
    \boldsymbol{C}(\ell) = 
    \begin{pmatrix}
     C^{\text{A}_0\text{A}_0}(\ell) & \dots & C^{\text{A}_0\text{A}_n}(\ell)\\
     \vdots & \ddots \\
     C^{\text{A}_n\text{A}_0}(\ell) & \dots & C^{\text{A}_n\text{A}_n}(\ell)
    \end{pmatrix}.
\end{equation}
Assuming that our observables $a_{\ell m}$ follow a multivariate Gaussian distribution, the expression for the Fisher matrix is:
\begin{equation}
\label{eqz:fisher}
    F_{\alpha\beta} = \sum_{\ell_{\text{min}}}^{\ell_{\text{max}}}\text{Tr}\left(\left(\boldsymbol{\Sigma}(\ell)\right)^{-1}\frac{\partial \boldsymbol{C}(\ell)}{\partial\theta_{\alpha}}\left(\boldsymbol{\Sigma}(\ell)\right)^{-1}\frac{\partial \boldsymbol{C}(\ell)}{\partial\theta_{\beta}}\right).
\end{equation}
\begin{table}
\footnotesize
\centering
\renewcommand\arraystretch{1.0}
\caption{Marginalised $1-\sigma$ errors divided by fiducial values and the percentage difference ($\Delta\%$) compared to~\cite{Euclid_2020} values. In each probe block, the first row shows the errors for the pessimistic scenario, and the second for the optimistic one.}
\label{tab:3x2}
\begin{tabular}{l|c|c|c|c|c|c|c}
  \toprule                       
  Probe & $\Omega_m$ & $\Omega_b$ & $h$ & $n_s$ & $\sigma_8$ & $w_0$ & $w_a$ \\
  \midrule 
  \multicolumn{8}{c}{$\nu\Lambda$CDM (fixed $M_{\nu}$)}\\
\midrule
\multirow{2}{6em}{3$\times$2pt} & 0.00799 & 0.0484 & 0.0257 & 0.00843 & 0.00381  & - & - \\ 
                                & 0.00279 & 0.0432 & 0.0190 & 0.00354 & 0.00134  & - & - \\ 
\midrule
\multirow{2}{6em}{\% difference} & 1.37 & 7.39 & 4.93 & 0.77 & 0.24 & - & - \\ 
                                 & 0.30 & 6.44 & 5.47 & 1.68 & 3.34 & - & - \\ 

\midrule

\multicolumn{8}{c}{$\nu w_0w_a$CDM (fixed $M_{\nu}$)}\\
\midrule
\multirow{2}{6em}{3$\times$2pt} & 0.0109 & 0.0517 & 0.0294 & 0.0104 & 0.00494 & 0.0420 & 0.172 \\ 
                                & 0.00607 & 0.0438 & 0.0191 & 0.00383 & 0.00232 & 0.0281 & 0.106 \\ 
\midrule
\multirow{2}{6em}{\% difference} & 0.87 & 4.38 & 1.50 & 3.96 & 2.83 & 0.04 & 1.37 \\ 
                                 & 2.86 & 5.11 & 4.94 & 1.81 & 5.05 & 4.00 & 5.98 \\ 

  \bottomrule                    
\end{tabular}
\end{table}
In order to compute the Fisher matrix, evaluation of the derivatives of the $C(\ell)$ with respect to cosmological and nuisance parameters is needed. The SteM fitting procedure~\cite{Camera_2016} has been applied which is based on a iterative linear regression: for each combination of probes and bin pair, A$_i-$B$_j$, multipole $\ell$, and cosmological parameter $\theta_{\alpha}$, we sample the $\theta_{\alpha}$-line in 15 points around $\theta_{\text{ref}}$ (this included), where the values of $\theta_{\text{ref}}$ are the ones in table~\ref{tab:fiducial}. More precisely, we take $\delta\theta_{\alpha} = 0$, $\pm 0.625\%$, $\pm 1.25\%$, $\pm 1.875\%$, $\pm 2.5\%$, $\pm3.75\%$, $\pm5\%$ and $\pm10\%$. By assuming that the neighbourhood is small enough around $\theta_{\text{ref}}$, all the $C^{\text{A}_i \text{B}_j}_{\theta_{\alpha}}(\ell)$ thus obtained should lie on a straight line. If the spread between the linearly fitted [$C^{\text{A}_i\text{B}_j}_{\theta_{\alpha}}(\ell)$]$^{\text{fit}}$ and the true values [$C^{\text{A}_i \text{B}_j}_{\theta_{\alpha}}(\ell)$]$^{\text{true}}$  is less than 1\%, we zoom in on the sampled $\theta_{\alpha}$-range by cutting out a few values on the edges, until we reach the requested accuracy. For each given combination $\{$A$_i, $B$_j , \ell, \theta_{\alpha}\}$, the numerical derivative $\partial C^{\text{A}_i \text{B}_j}(\ell)/\partial\theta_{\alpha}$ is the slope of the linear interpolation.\\
We checked the validity of our implementation confronting the results of the combination of the two probe, plus their cross-correlation CS+GC$_{\text{ph}}$+XC (usually known as $3\times2$pt) with~\cite{Euclid_2020}. The values of the marginalised $1-\sigma$ errors for both $\nu\Lambda$CDM and $\nu w_0w_a$CDM models with fixed $M_{\nu}$ are reported in the first row of table~\ref{tab:3x2} while in the second row the percentage difference between the two implementations. The contour plots for $\nu w_0w_a$CDM in the optimistic and pessimistic scenarios are shown in figure~\ref{fig:contour_3x2}.

\section{Stage-IV galaxy surveys and GW observables with the same \texorpdfstring{\boldmath}{}\texorpdfstring{$\ell_{\text{max}}$}{ell}}
\label{apndx_B}
We compare the Stage-IV galaxy surveys and GW probes assuming that they cover the same multipole range, i.e. $\ell_{\rm max}=1000$ and $\ell_{\rm max}=100$ for the optimistic and pessimistic case, respectively. We implement such a comparison with the aim to understand at a theoretical level (rather than an experimental one) the reason why the GW magnification looks to have such a poor constraining power with respect to Stage-IV, in particular to the CS. The values of the marginalised $1-\sigma$ errors are reported in the first part of table~\ref{tab:errori_3x2s_lcdm}. Here and in the following we define $3\times2{\rm pt}_{\ell_{\rm max}}$ as CS+GC$_{\rm ph}$+XC with all the $C(\ell)$ evaluated up to the same $\ell_{\rm max}$.

\subsection{The \texorpdfstring{$\nu\Lambda$CDM}{} scenario}
In this section we investigate the ability of the Stage-IV galaxy surveys and GW detectors combination in constraining the cosmological parameters of the $\nu\Lambda$CDM model, considering the total neutrino mass fixed. In particular we focus on the improvement due to the GW magnification with respect to the $3\times2{\rm pt}_{\ell_{\rm max}}$ statistics.

\begin{figure}[ht]
\centering
\setlength{\tabcolsep}{0.01pt}
\begin{tabular}{cc}
    \includegraphics[width=0.52\textwidth]{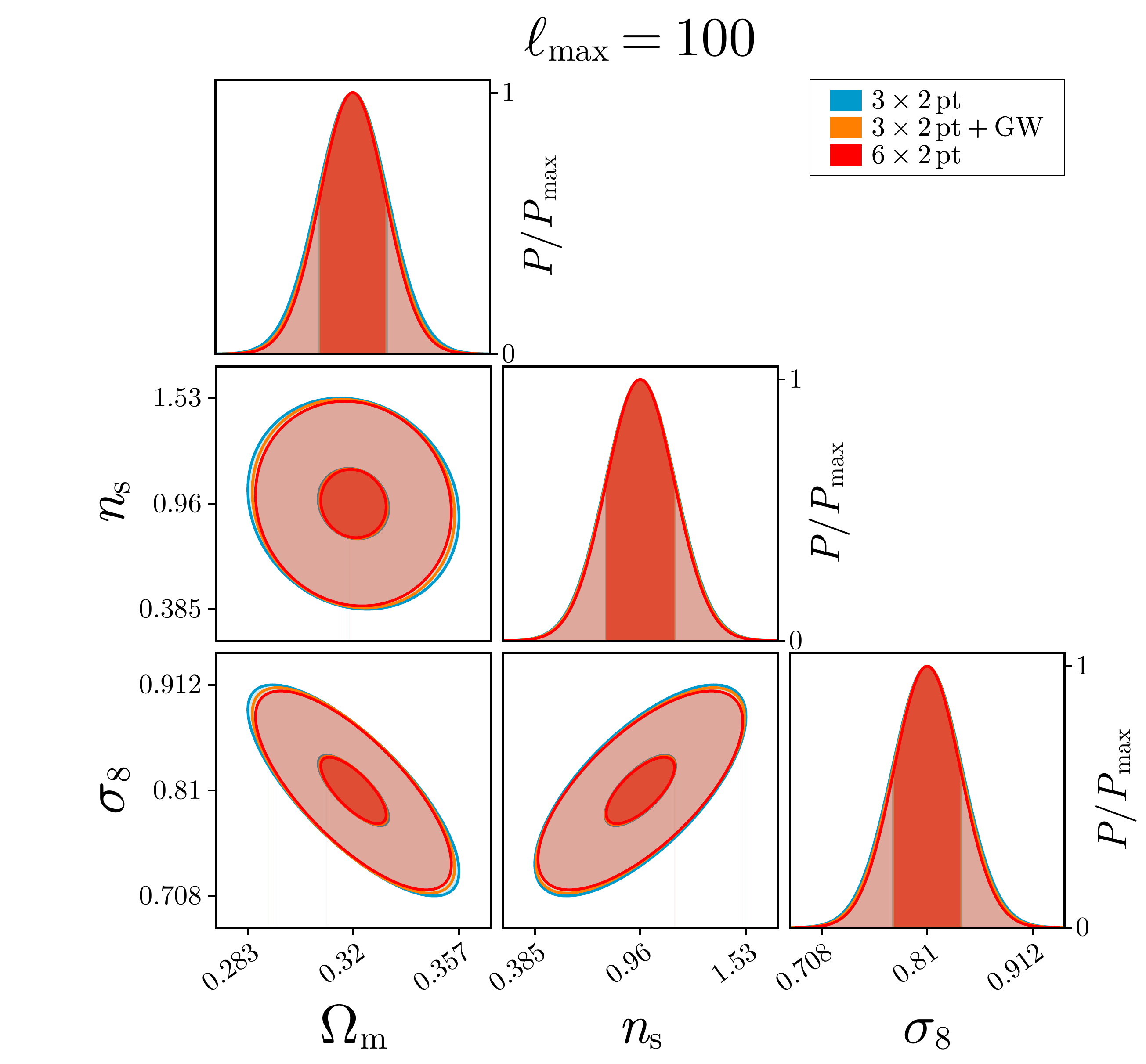} & \includegraphics[width=0.52\textwidth]{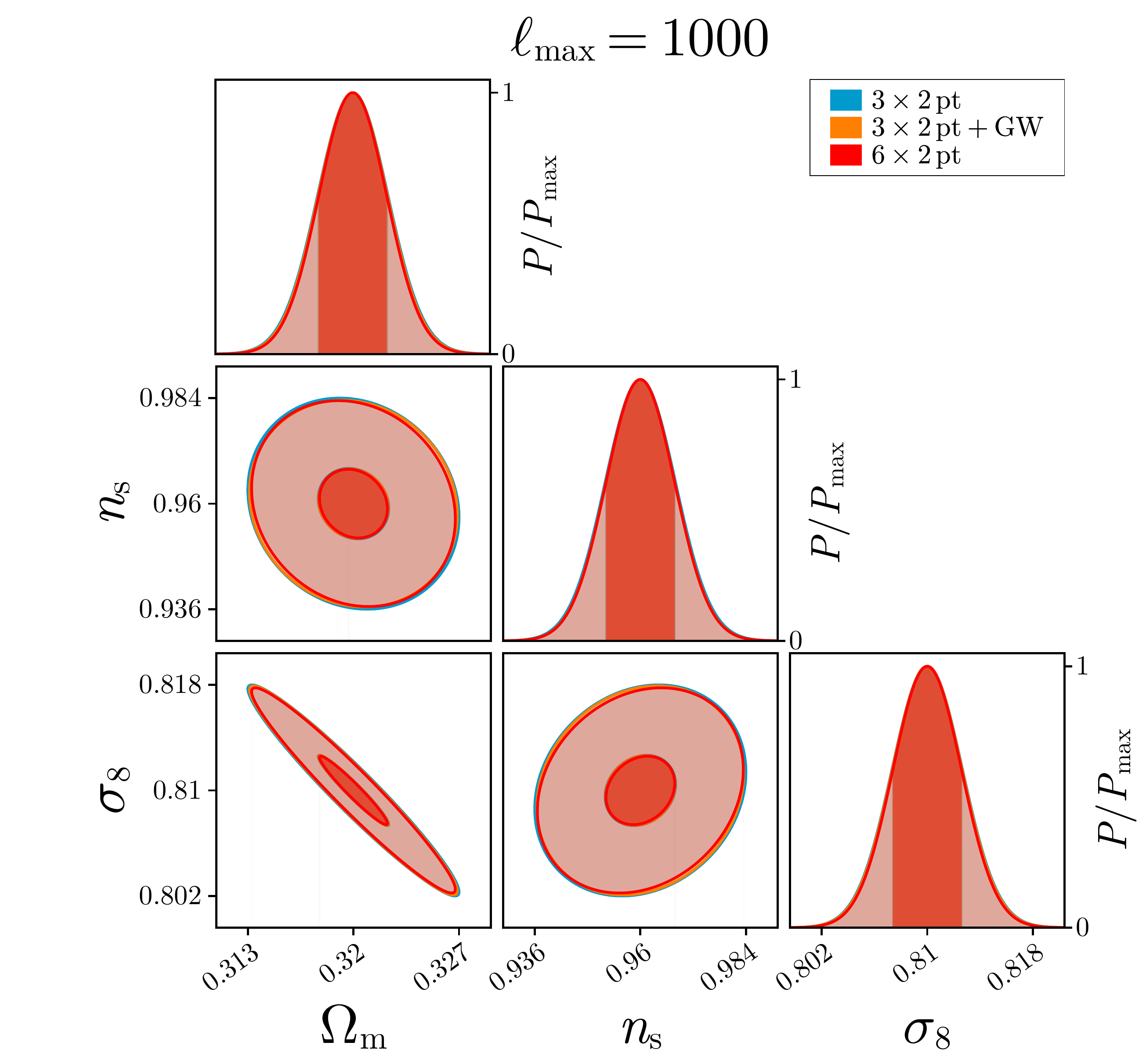} 
\end{tabular}
\caption{Fisher matrix marginalised contours for the $\nu\Lambda$CDM model with fixed neutrino mass limited to the multipole of the GW magnification signal, in the optimistic (right) and in the pessimistic (left) scenarios with $\ell_{\text{max}}=1000$ and $\ell_{\text{max}}=100$, respectively. Different scales between the the left and right panels have been used for visibility reasons.}
\label{fig:contour_6x2_ell_trunk_lcdm}
\end{figure}

When limiting the Stage-IV probes up to the same $\ell_{\rm max}$ of the GW signal the contribution of the GW magnification has a larger impact on parameter constraints when combined with CS and GC$_{\rm ph}$. In fact, there is a gain not only in the errors from the $6\times2$pt statistics, but also when considering the $3\times2$pt$_{\ell_{\rm max}}$ and GW magnification as independent probes. Concerning the $6\times2$pt statistics we have found that, in the pessimistic scenario, all the parameter constraints improve more than 2\%, reaching 7.26\% and 5.76\% in the case of $\Omega_m$ and $\sigma_8$, respectively when compared to the $3\times2{\rm pt}_{\ell_{\rm max}}$. In the optimistic case instead, the gain is a bit lower (reaching only 0.41\% and 1.16\% for $\Omega_b$ and $h$, respectively), meaning that the constraining power of GW is less dominant with respect to the other probes when a larger value of $\ell$ is adopted in the analysis (see table~\ref{tab:errori_lcdm}). Figure~\ref{fig:contour_6x2_ell_trunk_lcdm} shows the contour plots for the more affected parameters. Regarding the error improvements related to the inclusion of the GW probe only, with respect to the $3\times2{\rm pt}_{\ell_{\rm max}}$ in the pessimistic scenario, are $\sim$ 4\% for $\Omega_m$, $\sim$ 2.7\% for $\sigma_8$ and around 1.3\% for the other parameters. In the optimistic case, instead, the improvements are $\sim$ 4\% for $n_s$, and around 1\% for the remaining parameters.
\begin{figure}[ht]
\centering
\setlength{\tabcolsep}{0.01pt}
\begin{tabular}{cc}
\includegraphics[width=0.52\textwidth]{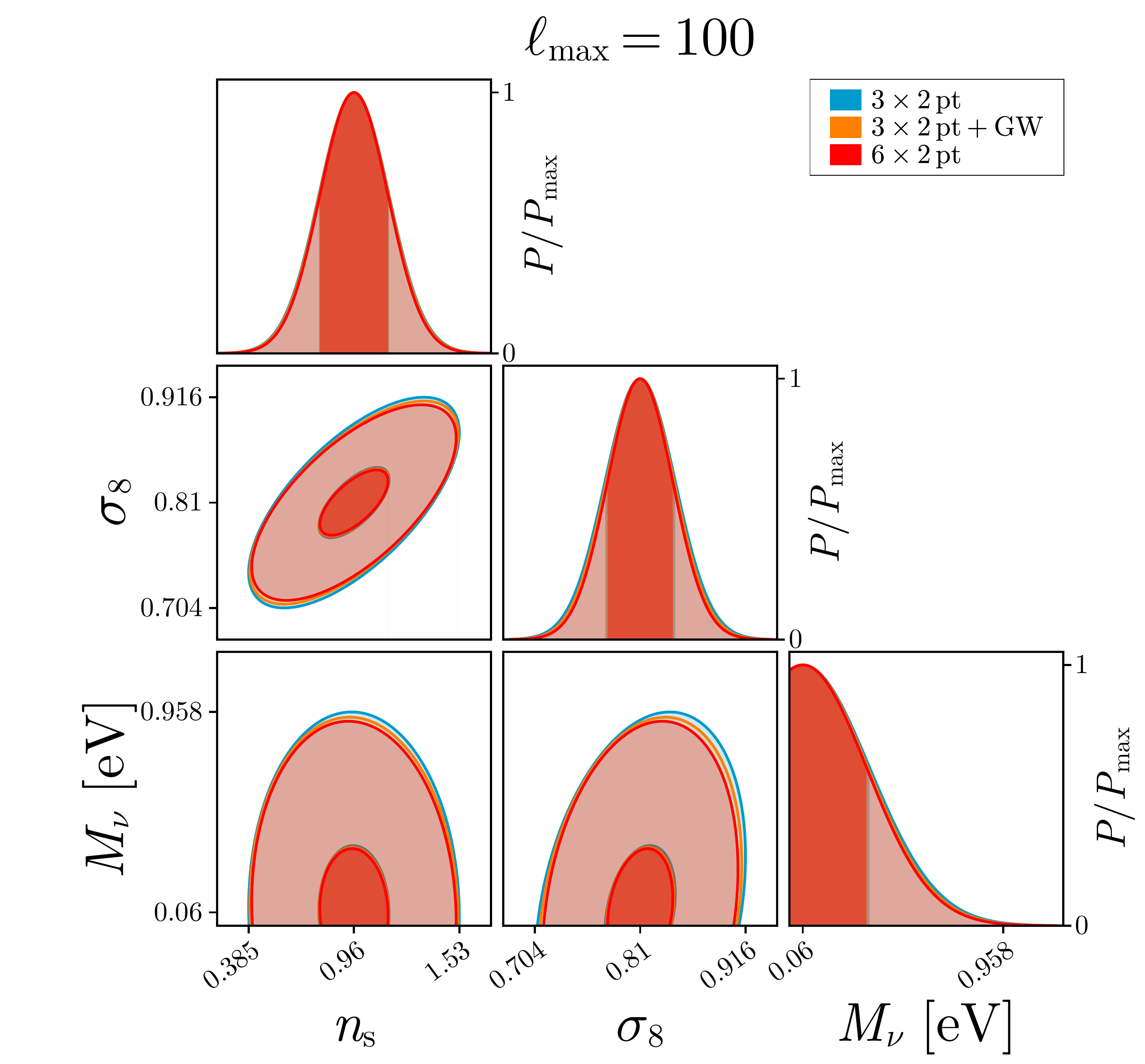} & \includegraphics[width=0.52\textwidth]{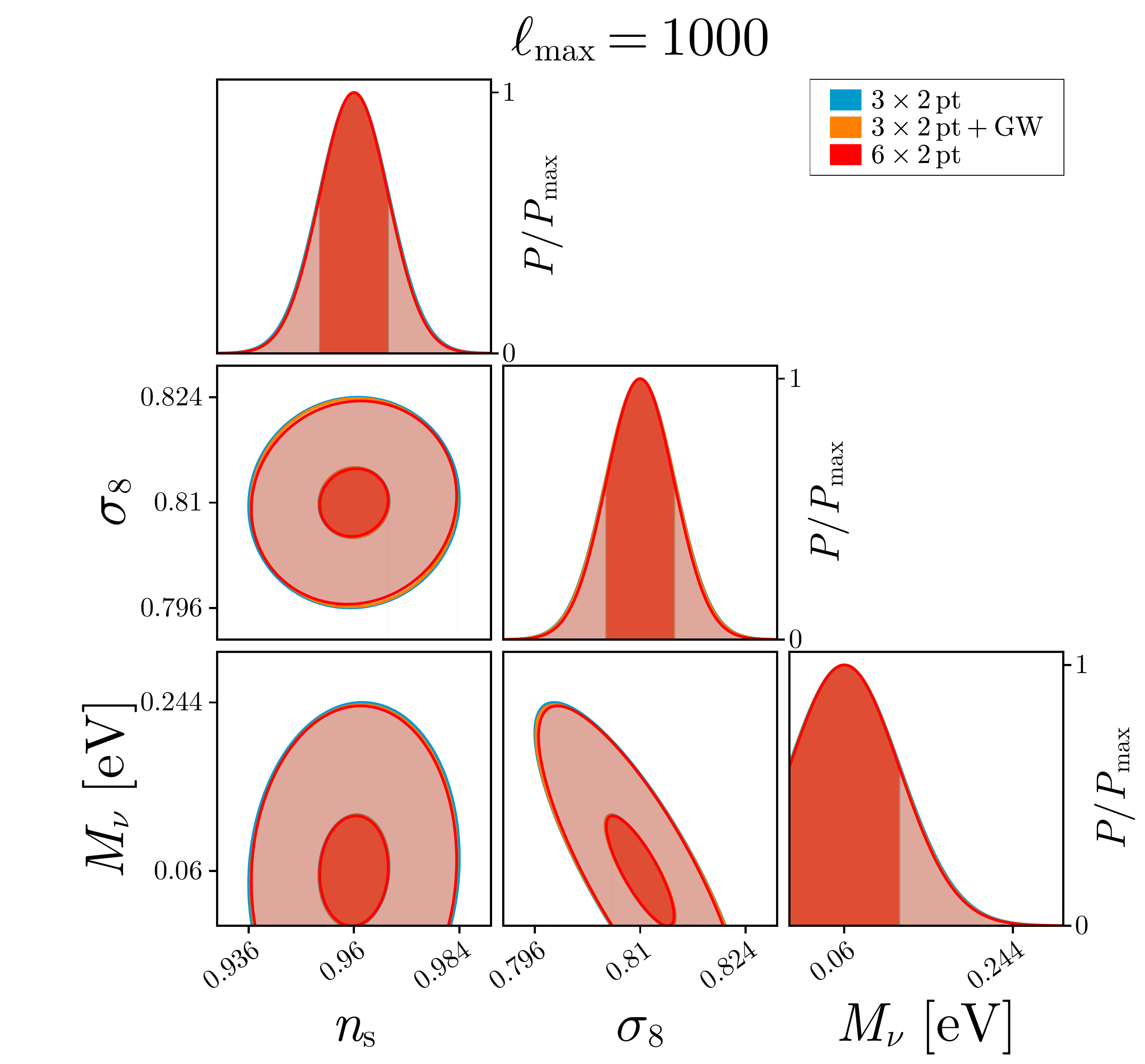} 
\end{tabular}
\caption{Fisher matrix marginalised contours for the $\nu\Lambda$CDM model limited to the multipole of the GW magnification signal, in the optimistic (right) and in the pessimistic (left) scenarios with $\ell_{\text{max}}=1000$ and $\ell_{\text{max}}=100$, respectively. Different scales between the
left and right panels have been used for visibility reasons.}
\label{fig:contour_6x2_ell_trunk_nulcdm}
\end{figure}

\subsection{The \texorpdfstring{$\nu\Lambda$CDM}{} scenario with free \texorpdfstring{$M_{\nu}$}{}}
Similarly to the $M_\nu$ fixed case, by limiting the analysis for all the probes to $\ell_{\rm max}=1000$, in the optimistic case, and to $\ell_{\rm max}=100$, in the pessimistic one, the GW probe has a much greater impact in improving the parameter constraints when combined with the $3\times2{\rm pt}_{\ell_{\rm max}}$ statistics.

From the second part of table~\ref{tab:errori_3x2s_lcdm} one can infer that the trend of the parameter constraints is maintained with respect to the $M_\nu$ fixed case, confirming that when all the probes span the same range of multipoles the imprint of the GW magnification probe on the parameter erros is more significant, especially in the pessimistic case where $\ell_{\rm max}=100$. In particular, the largest improvements can be observed for the errors on $\Omega_m$ and $\sigma_8$, while for the other parameters the improvement is similar to the case with fixed $M_\nu$. The trend of the parameter constraints, related to the addition of GW only, i.e neglecting its cross-correlation with the $3\times2{\rm pt}_{\ell_{\rm max}}$ statistics, is also maintained, but the percentage differences, with respect to the $3\times2{\rm pt}_{\ell_{\rm max}}$ statistics, of the errors are the half of  those for the $6\times2$pt statistics. Figure~\ref{fig:contour_6x2_ell_trunk_nulcdm} shows the contour plots for the most affected parameters in the considered scenario.
\begin{table}
\footnotesize
\centering
\renewcommand\arraystretch{1.0}
\caption{marginalised 1$-\sigma$ errors in the case all the probes share the same multipole range of the GW magnification signal. In each probe block, the first row shows the errors for the pessimistic scenario ($\ell_{\text{max}}=100$), and the second for the optimistic one ($\ell_{\text{max}}=1000$). The last row in each model block ($\Delta\%$) represents the percentage
difference between the $1-\sigma$ errors from the Stage-IV $3\times2$pt$_{\ell_{\rm max}}$ and the $6\times2$pt statistics. Here $3\times2{\rm pt}_{\ell_{\rm max}}\equiv$ CS+GC$_{\rm ph}$+XC with all the $C(\ell)$ evaluated at the same $\ell_{\rm max}$.}
\label{tab:errori_3x2s_lcdm}
\begin{tabular}{l|c|c|c|c|c|c|c|c}
  \toprule                       
  Probe & $\Omega_m$ & $\Omega_b$  & $h$ & $n_s$ & $\sigma_8$ & $M_{\nu}$[eV] & $w_0$ & $w_a$ \\
\midrule
\multicolumn{8}{c}{$\nu\Lambda$CDM (fixed $M_{\nu}$)}\\

\midrule
\multirow{2}{7em}{3$\times$2pt$_{\ell_{\rm max}}$} & 0.0124 & 0.0218 & 0.272 & 0.192 & 0.0339 & - & - & - \\ 
                                    & 0.00217 & 0.00232 & 0.0164 & 0.00797 & 0.00264 & - & - & - \\ 
\midrule
\multirow{2}{7.4em}{GW+3$\times$2pt$_{\ell_{\rm max}}$} & 0.0119 & 0.0216 & 0.270 & 0.189 & 0.0330 & - & - & - \\ 
                            & 0.00214 & 0.00232& 0.0164 & 0.00780 & 0.00261 & -  & - & - \\ 

\midrule
\multirow{2}{6em}{6$\times$2pt} & 0.0115 & 0.0212 & 0.266 & 0.186 & 0.0312 & - & - & - \\ 
                            & 0.00210 & 0.00232 & 0.0163 & 0.00777 & 0.00257 & - & - & - \\ 
\midrule
\multirow{2}{5em}{$\Delta$\%} & 7.26 & 2.69 & 2.46 & 3.06 & 5.76 & - & - & - \\ 
                            & 3.52 & 0.41& 1.16 & 2.47 & 2.83 & - & - & - \\ 

  \midrule 
\multicolumn{8}{c}{$\nu\Lambda$CDM}\\ 
\midrule
\multirow{2}{7em}{3$\times$2pt$_{\ell_{\rm max}}$} & 0.0133 & 0.0221 & 0.288 & 0.192 & 0.0353 & 0.299 & - & - \\ 
                                    & 0.00266 & 0.00233 & 0.0165 & 0.00799 & 0.00460 & 0.0615 & - & - \\ 
\midrule
\multirow{2}{7.4em}{GW+3$\times$2pt$_{\ell_{\rm max}}$} & 0.0126 & 0.0219 & 0.286 & 0.189 & 0.0340 & 0.291 & - & - \\ 
                & 0.00263 & 0.00232 & 0.0165 & 0.00782 & 0.00455 & 0.0607 & - & - \\

\midrule
\multirow{2}{6em}{6$\times$2pt} & 0.0121 & 0.0216 & 0.282 & 0.186 & 0.0328 & 0.285 & - & - \\ 
                            & 0.00255 & 0.00232 & 0.0163 & 0.00779 & 0.00443 & 0.0602 & - & - \\
\midrule
\multirow{2}{5em}{$\Delta$\%} & 9.07 & 2.31 & 2.08 & 2.94 & 7.16 & 4.63 & - & - \\ 
                            & 4.44 & 0.42 & 1.02 & 2.57 & 3.52 & 2.06 & - & - \\

     \midrule 
\multicolumn{8}{c}{$\nu w_0w_a$CDM (fixed $M_{\nu}$)}\\
\midrule
\multirow{2}{7em}{3$\times$2pt$_{\ell_{\rm max}}$} & 0.0208 & 0.0239 & 0.301 & 0.214 & 0.0508 & - & 0.181 & 0.898 \\ 
                                    & 0.00306 & 0.00248 & 0.0190 & 0.0102 & 0.00347 & - & 0.0388 & 0.162 \\ 
\midrule
\multirow{2}{7.4em}{GW+3$\times$2pt$_{\ell_{\rm max}}$} & 0.0206 & 0.0239 & 0.299 & 0.213 & 0.0505 & - & 0.176 & 0.886 \\ 
                & 0.00304 & 0.00247 & 0.0190 & 0.0100 & 0.00346 & - & 0.0386 & 0.160 \\

\midrule
\multirow{2}{6em}{6$\times$2pt} & 0.0191 & 0.0232 & 0.292 & 0.206 & 0.0471 & - & 0.167 & 0.834 \\ 
                            & 0.00295 & 0.00246 & 0.0189 & 0.0100 & 0.00341 & - & 0.0368 & 0.153 \\ 
\midrule
\multirow{2}{5em}{$\Delta$\%} & 8.06 & 3.25 & 3.05 & 3.74 & 7.39 & - & 7.69 & 7.19 \\ 
                            & 3.58 & 0.70 & 0.75 & 1.97 & 1.77 & - & 5.39 & 4.79 \\ 
  \midrule 
\multicolumn{8}{c}{$\nu w_0w_a$CDM}\\

\midrule
\multirow{2}{7em}{3$\times$2pt$_{\ell_{\rm max}}$} & 0.0245 & 0.0246 & 0.319 & 0.214 & 0.0589 & 0.570 & 0.198 & 1.28 \\ 
                                    & 0.00391 & 0.00266 & 0.0205 & 0.0127 & 0.00739 & 0.0890 & 0.0393 & 0.184 \\ 
\midrule
\multirow{2}{7.4em}{GW+3$\times$2pt$_{\ell_{\rm max}}$} & 0.0242 & 0.0245 & 0.314 & 0.213 & 0.0582 & 0.548 & 0.189 & 1.23 \\ 
                & 0.00389 & 0.00265& 0.0204 & 0.0125 & 0.00732 & 0.0876  & 0.0390 & 0.183 \\

\midrule
\multirow{2}{6em}{6$\times$2pt} & 0.0223 & 0.0238& 0.308 & 0.206 & 0.0541 & 0.516  & 0.179 & 1.14 \\ 
                            & 0.00381 & 0.00263 & 0.0202 & 0.0124 & 0.00727 & 0.0874 & 0.0372 & 0.177 \\
\midrule
\multirow{2}{5em}{$\Delta$\%} & 8.78 & 3.27 & 3.32 & 3.74 & 8.05 & 9.49 & 9.23 & 10.4 \\ 
                            & 2.47 & 0.91 & 1.06 & 2.34 & 1.58 & 1.88 & 5.20 & 3.91 \\
  \bottomrule                    
\end{tabular}
\end{table}
\begin{figure}[ht]
\centering
\setlength{\tabcolsep}{0.01pt}
\begin{tabular}{cc}
    \includegraphics[width=0.52\textwidth]{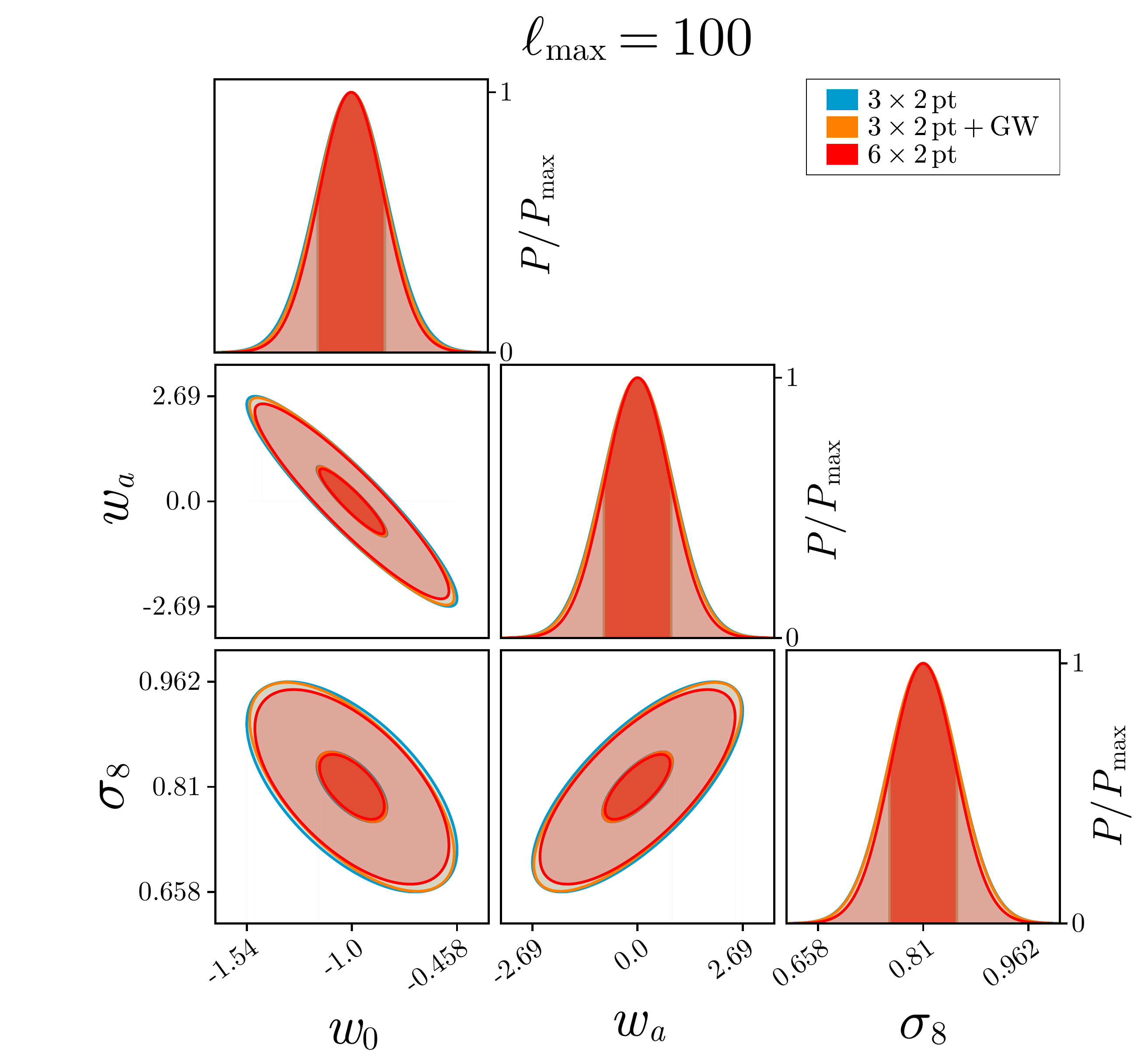} & \includegraphics[width=0.52\textwidth]{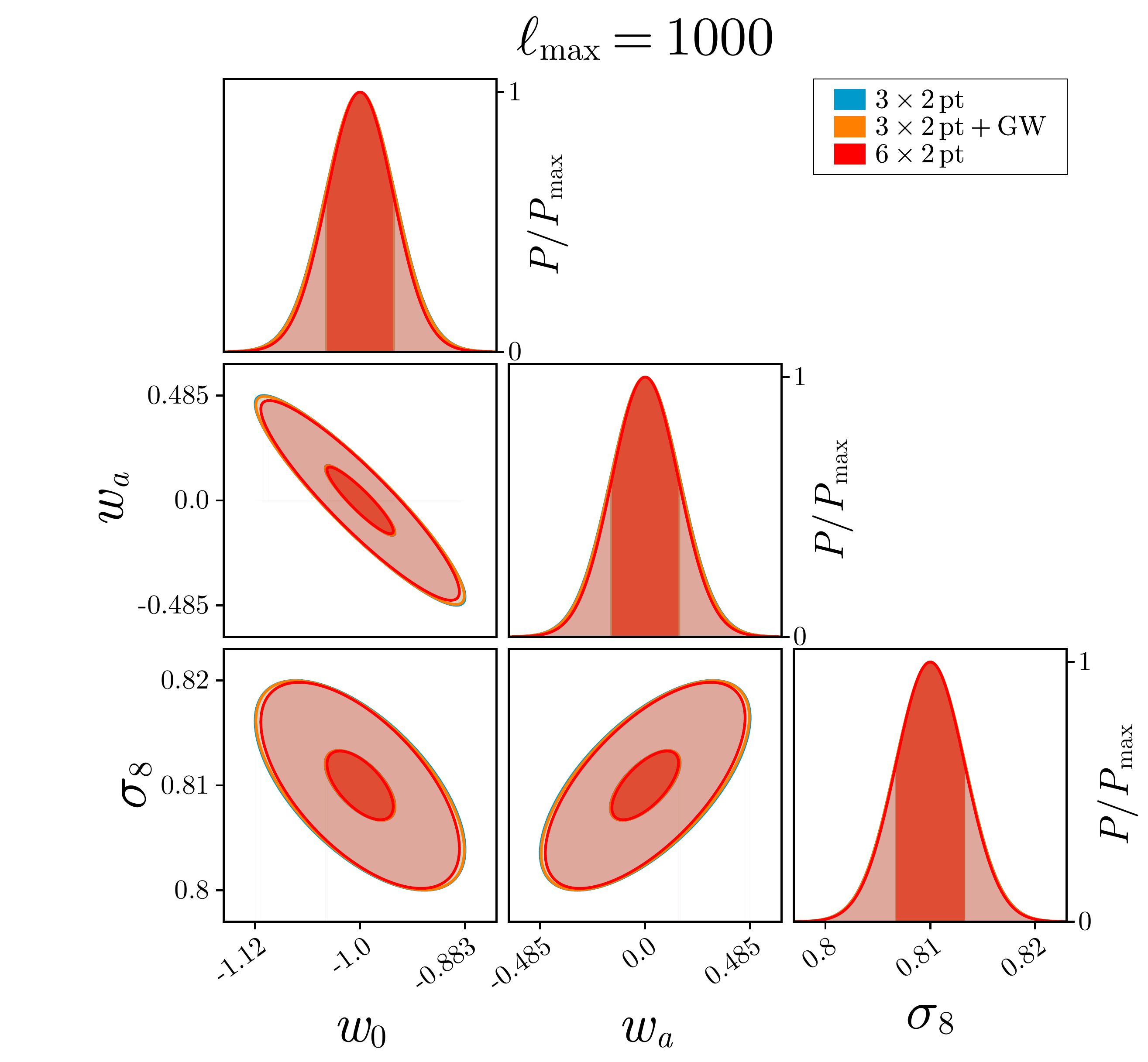} 
\end{tabular}
\caption{Fisher matrix marginalised contours for the $\nu w_0w_a$CDM with fixed neutrino mass limited to the multipole of the GW magnification signal, in the optimistic (right) and in the pessimistic (left) scenarios with $\ell_{\text{max}}=1000$ and $\ell_{\text{max}}=100$, respectively. Different scales between the
left and right panels have been used for visibility reasons.}
\label{fig:contour_6x2_ell_trunk_w0wa}
\end{figure}

\subsection{The \texorpdfstring{$\nu w_0w_a$CDM}{} scenario}
In this section we investigate the impact, on the Stage-IV galaxy surveys and GW detectors constraining power, of the DE equation of state parameters, $w_0$ and $w_a$, in the parameter space of the Fisher matrix analysis. Therefore, we focus on the differences of cosmological parameter errors with respect to the $\nu\Lambda$CDM case with fixed $M_\nu$, as well as on $w_0$ and $w_a$ constraints.

Again, when limiting the forecasts to the multipole range of the GW magnification signal,  one can immediately notice, from the third part of table~\ref{tab:errori_3x2s_lcdm}, the better improvement on the parameters constraints when exploiting the GW probe, especially in the pessimistic case of the $6\times2$pt statistics. However in this cosmological model the improvements, given by the addition of the GW probe, without the cross-correlation with CS and GC$_{\rm ph}$, are very small: the errors are almost similar to those given by the $3\times2{\rm pt}_{\ell_{\rm max}}$, except for $w_0$ and $w_a$ in the pessimistic case, whose errors improve by 2.8\% and 1.33\%. Figure~\ref{fig:contour_6x2_ell_trunk_w0wa} shows the contour plots for both the optimistic an pessimistic scenarios for the DE parameters together with $\sigma_8$.

\subsection{The \texorpdfstring{$\nu w_0w_a$CDM}{} scenario with free \texorpdfstring{$M_{\nu}$}{}}
\begin{figure}[ht]
\centering
\setlength{\tabcolsep}{0.01pt}
\begin{tabular}{cc}
\includegraphics[width=0.49\textwidth]{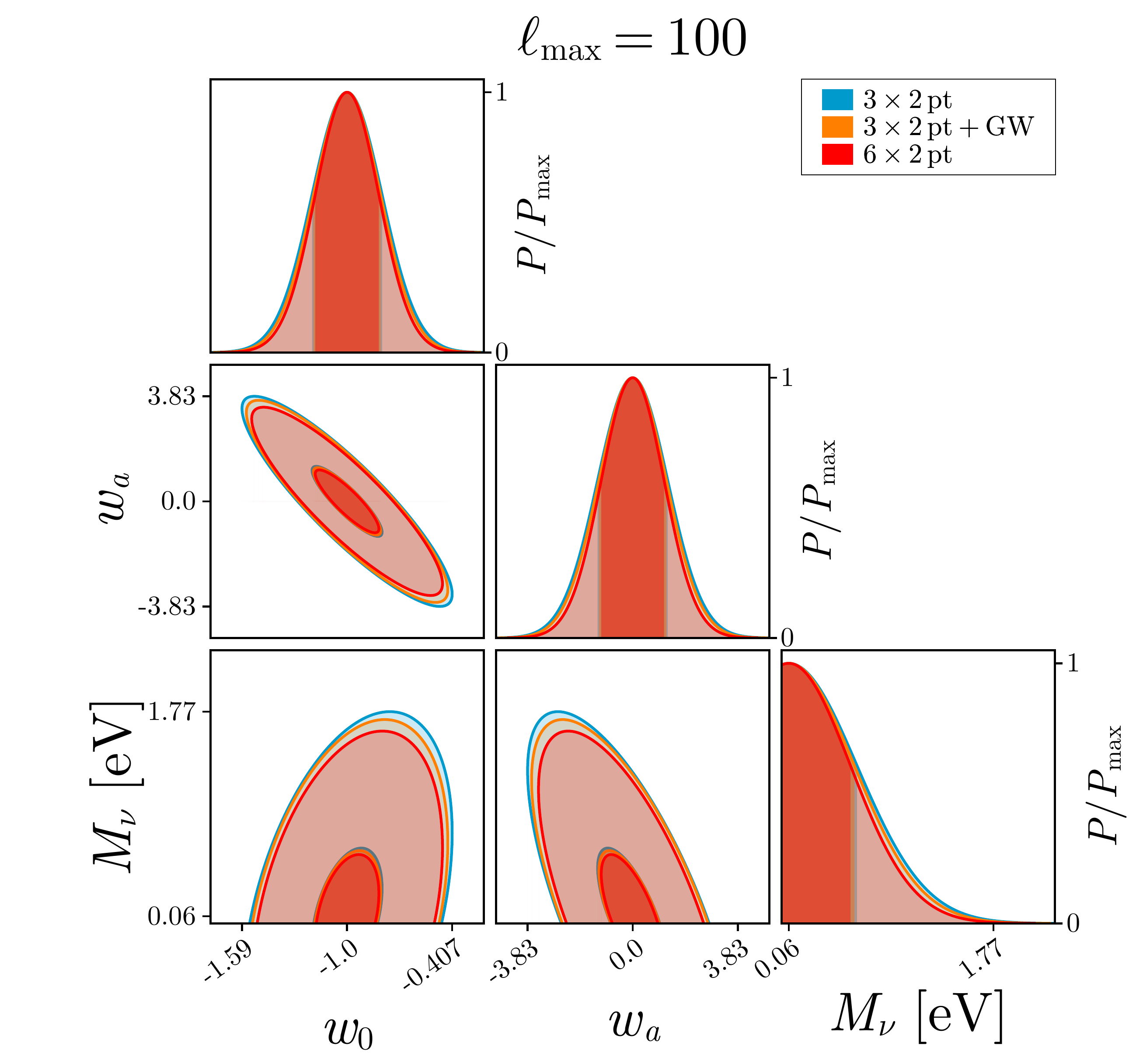} & \includegraphics[width=0.49\textwidth]{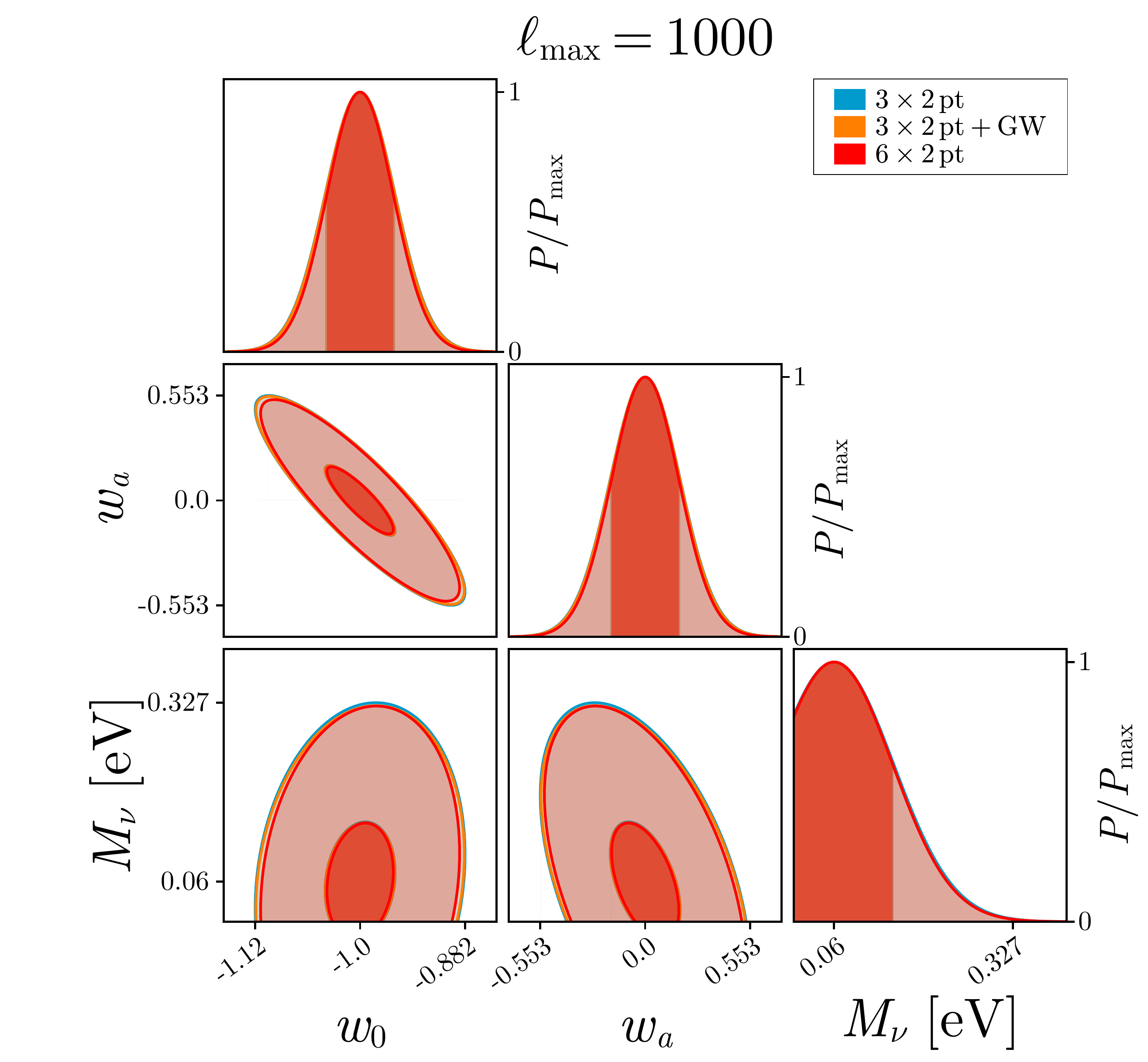} 
\end{tabular}
\caption{Fisher matrix marginalised contours for the $\nu w_0w_a$CDM limited to the multipole of the GW magnification signal, in the optimistic (right) and in the pessimistic (left) scenarios with $\ell_{\text{max}}=1000$ and $\ell_{\text{max}}=100$, respectively. Different scales between the left and right panels have been used for visibility reasons.}
\label{fig:contour_6x2_ell_trunk_nuw0wa}
\end{figure}

By limiting the multipoles to $\ell_{\rm max}=1000$ (100) for the optimistic (pessimistic) scenario, one obtains a larger value on the $1-\sigma$ errors, due to the smaller range of multipoles considered, but one can appreciate more the constraining power of GW that contribute in decreasing the errors on the cosmological parameters. In the last part table~\ref{tab:errori_3x2s_lcdm} we have reported the $1-\sigma$ errors, in both pessimistic (first row) and optimistic cases (second row), when adding the GW magnification probe, both as considered as an independent probe and when it is cross-correlated with the other two probes. Of the latter, the percentage improvement is reported and the contour plots for these combinations are shown in figure~\ref{fig:contour_6x2_ell_trunk_nuw0wa}. 

By looking at the constraints given by the addition of the GW, considered as an independent probe, to the $3\times2$pt, one can notice very tiny improvements in the error values (e.g. the most affected parameters, $w_a$ and $M_{\nu}$ gets a $\sim3.9$\% improvement). But if the cross-correlation between GW and the other two probes is accounted for, the constraints get tighter, providing improvements up to 10\% for the pessimistic case and up to 5\% in the optimistic one.
\section{Additional figures}
\begin{figure}[H]
  \centering
 \includegraphics[width=0.6\textwidth]{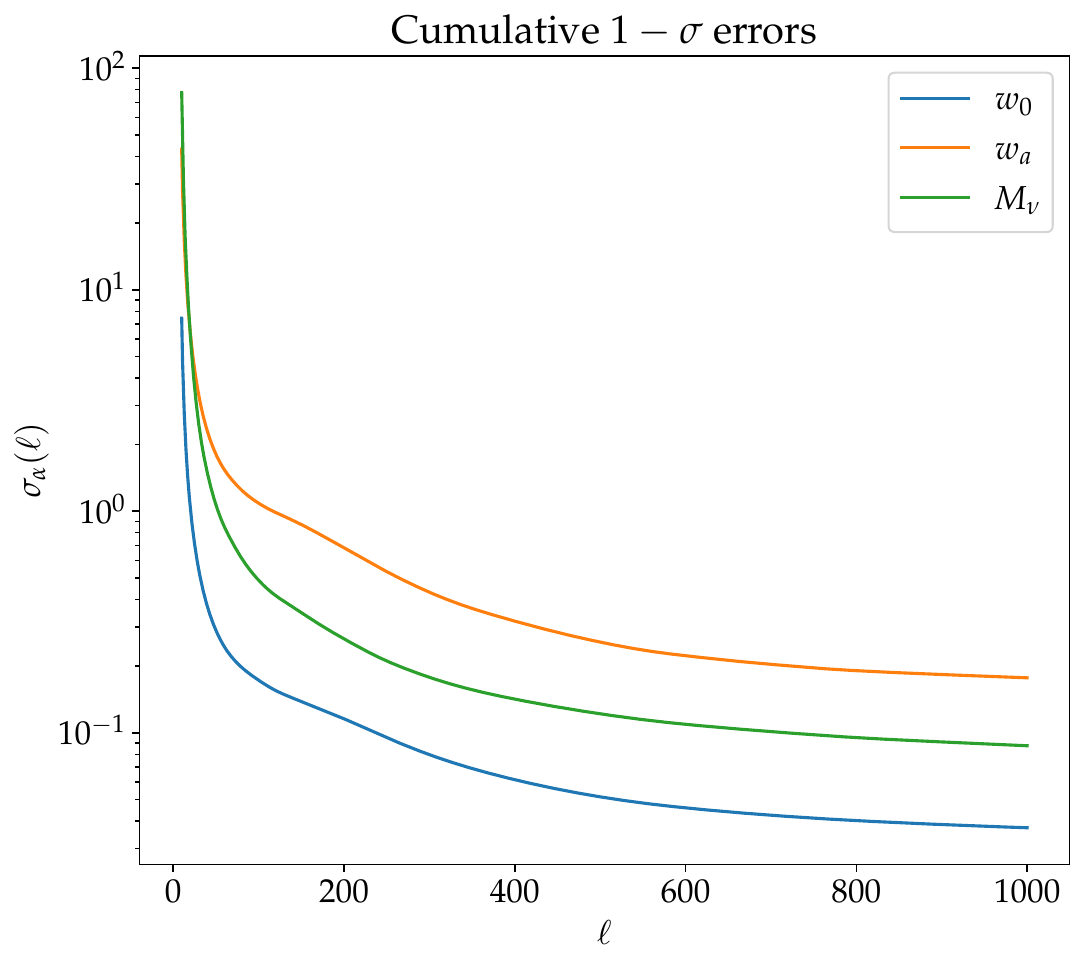}
  \caption{Cumulative marginalised $1-\sigma$ errors on various parameters. The values of the errors shown are the result of the summation of the $6\times2$pt Fisher matrices with increasing value of $\ell_{\rm max}$ with step $\Delta\ell=1$ in the interval $10\leq \ell_{\rm max} \leq 1000$.}
\label{fig:err_cum}
\end{figure}
\begin{figure}[H]
\begin{minipage}[b]{0.48\linewidth}
  \centering
 \includegraphics[width=\textwidth]{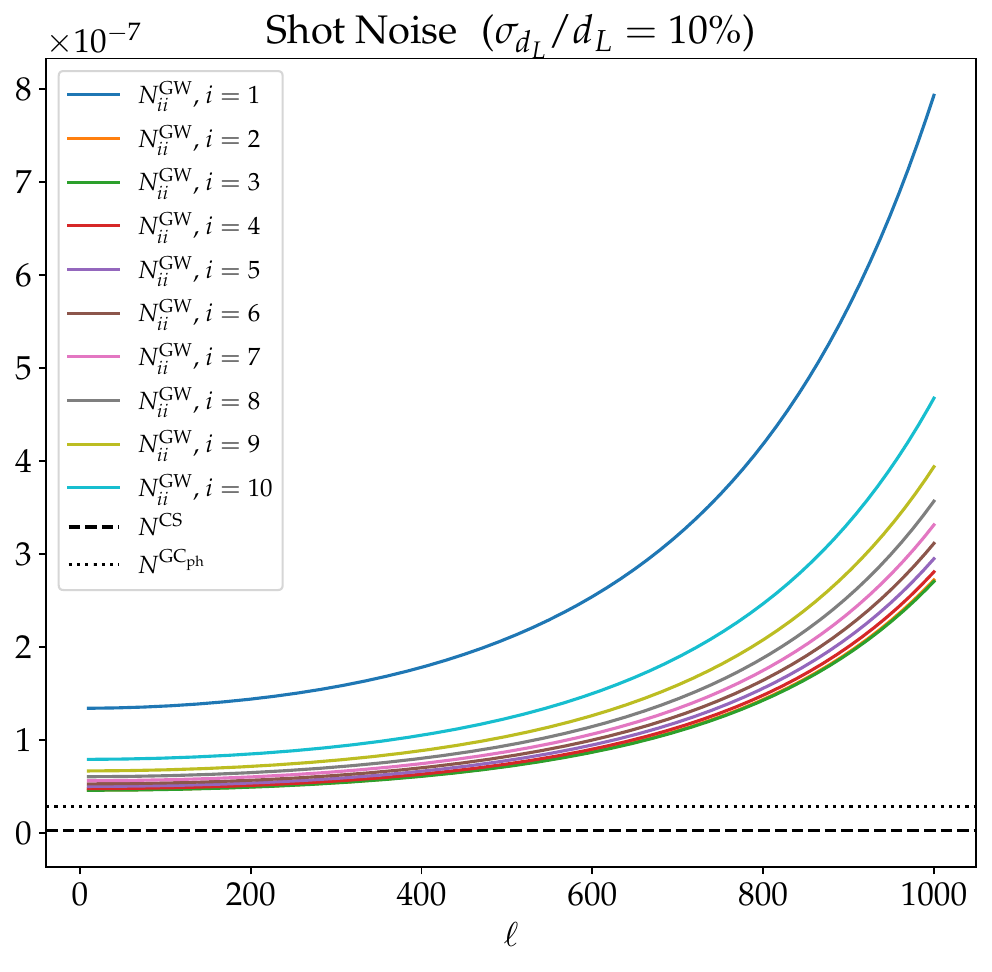}
\end{minipage}
\hfill
\begin{minipage}[b]{0.49\linewidth}
  \centering
   \includegraphics[width=\textwidth]{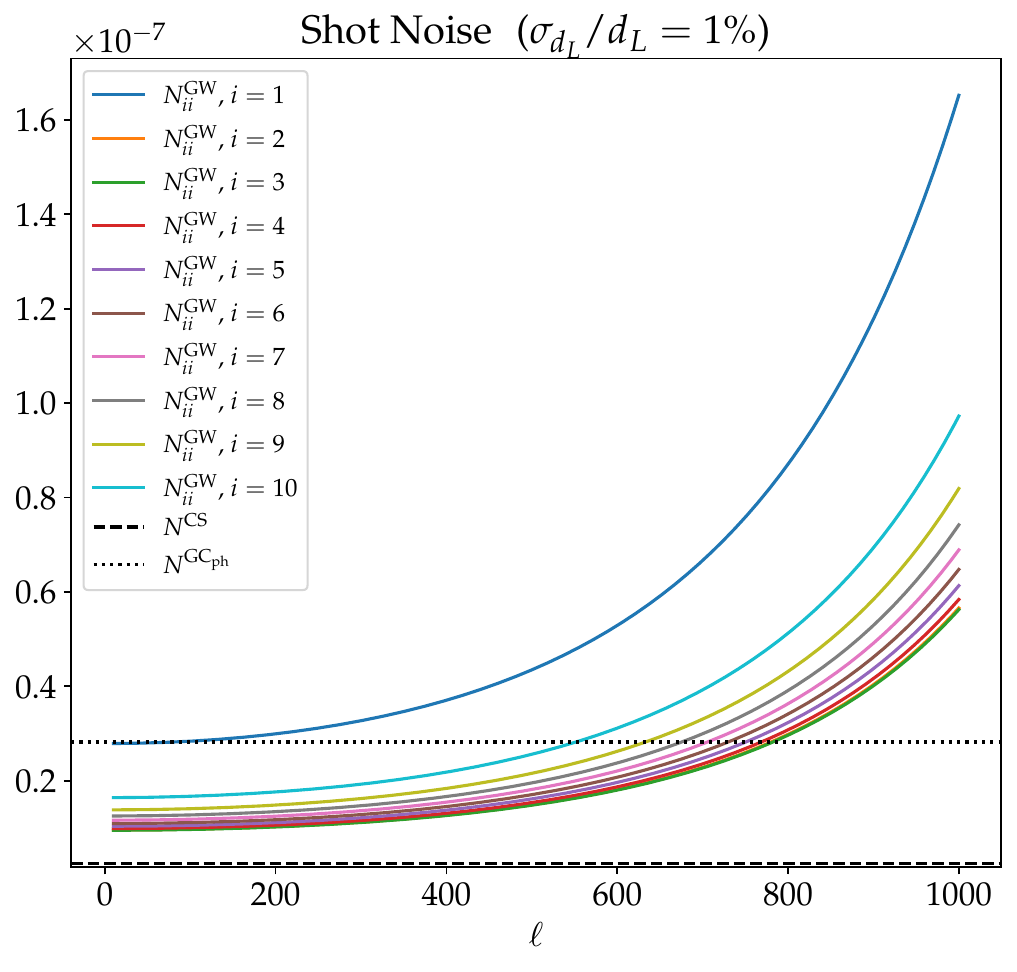}
\end{minipage}
\caption{Shot noise for the three different probes with different luminosity distance error.}
\label{fig:noise}
\end{figure}

\end{document}